\documentclass[12pt]{article}
\usepackage{amssymb}
\usepackage{amsfonts}
\usepackage{amsmath}
\usepackage{amsmath,amsthm}
\usepackage{subfigure}
\usepackage{graphicx,epsfig, color }

\oddsidemargin 10mm \numberwithin{equation}{section}
\begin{document}
\begin{center}
{{\bf {Thermodynamics of 4D dS/AdS Gauss-Bonnet black holes from
consistent gravity theory in presence of cloud of strings}} \vskip
0.5 cm
 Hossein Ghaffarnejad\footnote{E-mail address:
hghafarnejad@semnan.ac.ir}\vskip 0.5 cm
   \textit{Faculty of Physics, Semnan University, P.C. 35131-19111, Semnan,
   Iran}}\\
    \end{center}
\begin{abstract}
By looking at the Lovelock theorem one can infer that the gravity
model given by \cite{GL} could not applicable for all type of 4D
Einstein Gauss Bonnet (GB) curved spacetimes. Because in 4D space
time the Gauss-Bonnet invariant is a total derivative, and hence
it does not contribute to the gravitational dynamics. Hence,
authors in the work \cite{AL} presented an alternative consistent
EGB gravity model instead of \cite{GL} by applying break of
diffeomorphism property. In this work we use it to produce a
dS/AdS black hole metric and then investigate its thermodynamic
behavior in presence of cloud of Numbo Goto strings.
\end{abstract}
\section{Introduction}
Among the various higher order derivative gravitational models
which are given in the literature,  Lovelock gravity \cite{Lovl}
is quite special because of bing free of ghost
\cite{s1,s2,s3,s6,s7,s8,s9}. In fact, there are presented many
higher order derivative metric theories which show Ostrogradsky
instability (see \cite{woo, Haya} for a good review). In this
sense the actions which contains higher order curvature terms
introduce equations of motions with fourth order or higher metric
derivatives where linear perturbations disclose that the graviton
should be a ghost.  Fortunately the Lovelock model is free of
ghost term  and so have field equations involving not more than
second order derivatives of the metric.  Action functional of the
Lovelock gravity, is given by combinations of various terms as
follows. The first term is the cosmological constant $\Lambda,$
the second term is the Ricci scalar $R=R_\mu^\mu,$ and the third
and fourth terms are the second order Gauss-Bonnet \cite{Lan} and
third order Lovelock terms (see Eq. 22 in ref. \cite{Sus}),
respectively.
 Without the
latter term the Lovelock gravity reduces to a simplest form called
as the Einstein-Gauss-Bonnet (EGB) theory in which the
Einstein-Hilbert action is supplemented with the quadratic
curvature GB term as source of self interaction of gravity.
Importance of this form of the gravity model is appeared more when
we see that it is  generated from effective Lagrangian of low
energy string theory \cite{e1,e2,e3,e4,e5}. In fact for higher
than 4 dimensions of curved spacetimes the Gauss Bonnet coupling
parameter which is calculated by dimensional regularization method
have some regular values but not for $D=4$. To resolve this
problem the author Glaven and his collaborator presented a
proposal \cite{GL} but we know now that their initial proposal
does not lead to every well-defined gravity theory because
regularization is guaranteed  just with some of metric theories
but not for whole of them. In this sense one can see \cite{cano,
cano1} in which authors explained several inconsistencies of the
original paper given by Glavan and Lin \cite{GL}. Particularly,
besides pointing out possible problems in defining the limit or
finding an action for the theory, their work also adds new results
to the discussion concerning the indefiniteness of second order
perturbations even at a Minkowskian background and the geodesic
incompleteness of spherically symmetric black hole geometry
presented by Glavan and Lin (See also \cite{Bayram}). Thus there
must be presented other proposals that can be cover all metric
theories. In response to this problem recently a well defined and
consistent theory is presented \cite{AL} by breaking the
diffeomorphism property of the curved spacetime. Instead of the
former work \cite{GL} the latter model
 is in concordance with the Lovelock theorem and thereby, seems more to be physical and applicable. For instance FRW cosmology of the latter model
  is studied in \cite{AL2} and showed success of this model with
  respect to the work
  \cite{GL}.
  In fact there are many papers which are published
 in the literature about the 4DEGB gravity  and its applications in four or higher dimensions of the space time
  which one can see collections of these works addressed in introduction of the reference
  \cite{Ghaf}. Here we point just to some newest  works as
  follows.
  For instance one can see \cite{Tae} where the authors
obtained an exact static, spherically symmetric black hole
solution in presence of the third-order Lovelock gravity with a
string cloud background in seven dimensions when the second- and
third-order Lovelock coefficients are related via
$\alpha^2_2=3\alpha_3$. Further, they examined thermodynamic
properties of this black hole to obtain exact expressions for
mass, temperature, heat capacity and entropy, and also perform the
thermodynamic stability analysis. In their work we see that a
string cloud background has a profound influence on the horizon
structure, thermodynamic properties, and the stability of black
holes. Interestingly the entropy of the black hole is unaffected
due to the string cloud background. However, the critical solution
for thermodynamic stability is affected by the string cloud
background. Similar work was investigated by Toledo and his
collaborator \cite{Jeff} in presence of quintessence but for
different spacetime dimensions. They showed the graphs
corresponding to the mass and the Hawking temperature for
different dimensions of spacetime, $D=4,5,6,7$. By concerning
Hawking radiation, it is shown that the radiation spectrum is
related to the change of entropy which codifies the presence of
the cloud of strings as well as of the quintessence. In their work
importance of number of space time dimensions are shown in thermal
stabilization of the black holes with effect of string surrounded
with quintessence. By studying relation between the Hawking
temperature and the entropy they discussed that radiation rate and
showed that this quantity depends on the change of entropy which
is given in terms of the event horizon and it is strongly
influenced by the presence of the cloud of strings as well as of
the quintessence. Therefore, the Hawking radiation spectrum
depends strongly on the presence of the cloud of strings and on
the quintessence.  In these view one can infer that presence of
string clouds cause to be stable a black hole thermodynamically.
Regarding the importance of the role of string theory in the study
of black hole dynamics we know that Juan Maldacena (see for a good
review \cite{Mal}), explained for a first time that the
development of a string theoretic interpretation of black holes
where quantum mechanics and general relativity, theories
previously considered incompatible, are united. Work by Maldacena
and others has given a precise description of a black hole, which
is described holographically in terms of a theory living on the
horizon. According to this theory, black holes behave like
ordinary quantum mechanical objects-information about them is not
lost, as previously thought, but retained on their
horizons-leading physicists to look at black holes as laboratories
for describing the quantum mechanics of spacetime and for modeling
strongly interacting quantum systems. Furthermore the authors of
the work \cite{dharm} used model\cite{GL} to obtain EGB
spherically symmetric static charged black hole in presence of the
Maxwell EM fields and cloud of strings. They obtained that by
owing to the corrected black hole due to the background cloud of
string, the thermodynamic quantities have also been corrected
except for the entropy, which remains unaffected by a cloud of
string background. The Bekenstein-Hawking area law turns out to be
corrected by a logarithmic area term. The heat capacity diverges
to infinity at a critical radius where incidentally the
temperature has a maximum, and the Hawking-Page transitions even
in absence of the cosmological term is happened by allowing the
black hole to become thermodynamically stable. The smaller black
holes are globally preferred with negative free energy. Also their
solution can also be identified as a $4D$ monopole-charged EGB
black hole. Particularly their solution reaches to spherically
symmetric black hole solutions of general relativity
asymptotically in the limits
$\alpha \to 0$ and absence of the string tension.\\
   In this work we use the consistent EGB gravity
  model \cite{AL} in minisuperspace approach and obtain metric of a spherically symmetric static
  charge-less black
  hole in presence of a cosmological parameter and Numbo
  Goto string tension.
 Metric field equations are solved numerically in which we use Runge Kutta method to produce numeric
  values of the fields with best fitting functions.
 Then we investigated thermodynamic behavior of the obtained
solution. To do so we calculate equation of state generated by
Hawking temperature of the black hole solution. In fact in the
extended phase space the cosmological constant plays an important
as thermodynamic pressure of vacuum dS (AdS) Sitter background
space. In our obtained metric solutions we will see that the
Gauss-Bonnet coupling constant plays a critical role to determine
scale of the black hole and positions of the critical points in
phase space where the black hole can be participate in the small
to large black hole phase transition or Hawking-Page phase
transition. In the latter one an unstable black hole reaches to
the dS/AdS vacuum space finally.\\ Setup of this work is as
follows.  In section 2, we recall consistent 4DEGB gravity given
by \cite{AL} and use Numbo Goto string fluid as matter source of
the system under consideration. In section 3 we generate metric
field equations for spherically symmetric 4D black hole line
element. In section 4 we solve metric field equations without the
string tension where the cosmological constant is just the source.
In this case the field equations take on  simpler forms and so we
obtained analytic form for the metric fields. In order to
numerically solve the field equations in presence of the string
tension we provide some physical initial conditions in Section 5.
In section 6 we apply to numeric analysis of the solutions. Last
section denotes to concluding remark and outlook.
\section{4D dS/AdS GB gravity with string fluid}
According to the work \cite{AL} we know a consistent EGB gravity
theory in $D\to4$ limit is given by first  part of the following
action functional.
\begin{equation}\label{action}I=\frac{1}{16\pi G}\int
\textrm{d}t\textrm{d}^3xN\sqrt{\gamma}(\mathcal{L}_{\textrm{EGB}}^{\textrm{4D}}+
\mathcal{L}_{\mathrm{matter}}),
 \end{equation} in which the Einstein Gauss Bonnet geometrical  Lagrangian density is
\begin{equation} \label{lagEGB}
 \mathcal{L}_{\mathrm{EGB}}^{4D}=
 2R-2\Lambda-\mathcal{M}\end{equation}$$+\frac{\tilde{\alpha}}{2}[8R^2-4R\mathcal{M}-\mathcal{M}^2-\frac{8}{3}(8R_{ij}R^{ij}-
 4R_{ij}\mathcal{M}^{ij}
 -\mathcal{M}_{ij}\mathcal{M}^{ij})].
$$
and the second part is matter Lagrangian density. The parameter
$G$ is the Newton`s gravitational coupling constant,
$R=R^i_i(\gamma_{ij})$ and $R_{ij}(\gamma_{ij})$ are the Ricci
scalar and the Ricci tensor for the spatial metric $\gamma_{ij}$
respectively and
\begin{equation}\mathcal{M}_{ij}=R_{ij}+\mathcal{K}_k^k\mathcal{K}_{ij}-\mathcal{K}_{ik}\mathcal{K}^{k}_j,~~~~\mathcal{M}=\mathcal{M}_i^i
\end{equation}
with
\begin{equation}\mathcal{K}_{ij}=\frac{1}{2N}(\dot{\gamma}_{ij}-2D_iN_j-2D_jN_i-\gamma_{ij}D_kD^k
\lambda_{\textrm{GF}}).\end{equation} Here a dot denotes time
derivative $t$ and all the effects of the constraint stemming from
the gauge-fixing (GF) are now encoded in Lagrange multiplier
$\lambda_{GF}$. $D_i$ is spatial covariant derivative and
re-scaled regular GB coupling constant $\tilde{\alpha}$ is defined
versus the irregular GB coupling constant $\alpha_{GB}$ in limits
of $D\to4$ dimensions such that $\tilde{\alpha}=(D-4)\alpha_{GB}.$
The above EGB gravity action satisfies the following gauge
condition for all spherically symmetric and cosmological
backgrounds (see \cite{AL} and \cite{AL2}).
\begin{equation}\sqrt{\gamma}D_kD^k(\pi^{ij}\gamma_{ij}/\sqrt{\gamma})\approx0.\end{equation}
In fact the above EGB action is rewritten versus the 1+3 ADM
decomposition of the a 4D background metric for which
\begin{equation}\label{met1}ds^2=g_{\mu\nu}dx^\mu dx^\nu=-N^2dt^2+\gamma_{ij}(dx^i+N^idt)(dx^j+N^jdt)\end{equation}
where $N, N_i, \gamma_{ij}$ are the lapse function, the shift
vector, and the spatial metric respectively. $\gamma$ factor in
the action (\ref{action}) is absolute value of determinant of the
spatial metric $\gamma_{ij}.$ This ADM decomposition is done on
the background metric to remove divergent boundary terms of the
higher order metric derivative in the GB term of the action
functional (\ref{action}) in general 4D form \cite{AL}. First term
in the  theory  defined  by (\ref{action}) has the time
re-parametrization symmetry $t\to t=t(t^\prime).$ We set matter
Lagrangian density $\mathcal{L}_{\mathrm{matter}}$ to be the
Nambu-Goto \cite{Let} (see also page 100 in ref. \cite{Bar}) which
explains dynamics of relativistic strings as follows.
\begin{equation}\label{NG}
 I_{\mathrm{\textrm{NG}}}=-\int_{\Sigma} \mathcal{L}_{\textrm{NG}} \textrm{d}\sigma^{0} \textrm{d}\sigma^{1}
\end{equation}
in which
\begin{equation}\label{LNG}\mathcal{L}_{\textrm{NG}}=\rho\sqrt{\mathfrak{g}}=\rho\big(-\frac{1}{2}
\Sigma^{\mu\nu}\Sigma_{\mu\nu}\big)^\frac{1}{2}\end{equation}
is lagrangian density and the constant parameter $\rho$ is tension
or proper density of the string. Also the bivector
$\Sigma^{\mu\nu}$  is related to string worldsheet parameters
$(\sigma^0,\sigma^1)$ such that
\begin{equation} \label{bi}\Sigma^{\mu \nu}=\epsilon^{ab}
\frac{\partial x^{\mu}}{\partial\sigma^{a}} \frac{\partial
x^{\nu}}{\partial \sigma^{b}}
\end{equation}
for which $\epsilon^{ab}$ is two dimensional Levi-Civita tensor
density $\epsilon^{01}=-\epsilon^{10}=1$ and $\mathfrak{g}$ is
absolute value of determinant of induced metric
\begin{equation}\label{met3}
\mathfrak{g}_{a b} = g_{\mu \nu} \frac{\partial x^{\mu}}{\partial
\sigma^{a}} \frac{\partial x^{\nu}}{\partial \sigma^{b}}.
\end{equation}
Stress energy tensor of this relativistic string is given by
\begin{equation}\label{stress}
T^{\mu \nu}=\mathfrak{g}^{-\frac{1}{2}}\rho \Sigma^{\mu
\delta}\Sigma_{\delta}^{\nu}=\frac{2\partial\mathfrak{L}_{NG}}{\partial
g^{\mu\nu}}
\end{equation}
 and its covariant conservation reads $\nabla_{\mu}(\rho\Sigma^{\mu\nu})=0.$ To have a $T_{\mu\nu}$ invariant under a re-parametrization
 of the string`s world sheets $\rho$ must be transformed as $\mathfrak{g}^{-1/2}$ (see \cite{Let} and references therein). Now we are in
 position to use the above model for a spherically symmetric static black hole spacetime.
 \section{4D dS/AdS GBBH surrounded by string cloud}
  By comparing the line element (\ref{met1}) with general anisotropic form of spherically symmetric static 4D curved
  metric
\begin{equation}\label{line1}ds^2=-e^{2A(r)}\bigg(1-\frac{2M(r)}{r}\bigg)dt^2+\frac{dr^2}{1-\frac{2M(r)}{r}}+r^2d\theta^2+r^2\sin^2\theta
d\varphi^2
 \end{equation} we obtain
\begin{equation}\label{def} N=e^{A(r)}\sqrt{1-\frac{2M(r)}{r}},~~~N_{r,\theta,\varphi}=0,\end{equation}$$\gamma_{rr}=\frac{1}{1-\frac{2M(r)}{r}}~
 ~~\gamma_{\theta\theta}=r^2
 ,~
 ~~\gamma_{\varphi\varphi}=r^2\sin^2\theta,~~~\lambda_{\textrm{GF}}=\lambda_{\textrm{GF}}(r).$$
By substituting (\ref{line1})
 into (\ref{lagEGB}) we obtain
\begin{equation}\label{EGB}\mathcal{L}^{\textrm{4D}}_{\textrm{EGB}}=R(\gamma_{ij})-2\Lambda+12q^2+
\frac{\tilde{\alpha}}{2}\bigg[3R^2(\gamma)+\frac{88}{3}q^2R(\gamma)
-272q^4-8R_{ij}(\gamma)R^{ij}(\gamma)\bigg]\end{equation} in which
\begin{equation}\label{q}q(r)=\frac{e^{-A(r)}}{r^2}\bigg[r^2\lambda^\prime_{GF}(r)\bigg]^\prime,
~~~\mathcal{K}_{ij}=-q\gamma_{ij}
\end{equation}
 \begin{equation}\label{R}R(\gamma)=-\frac{4M^\prime}{r^2},~~~R_{ij}(\gamma)R^{ij}(\gamma)=\frac{6}{r^6}(M-rM^\prime)^2\end{equation}
 and $\prime$ denotes derivative with respect to $r$.
 To obtain explicit form of the lagrangian
density of the NG string for the line element (\ref{line1}) we
should solve $\nabla_{\mu}(\rho\Sigma^{\mu\nu})=0.$ But this
equation is equivalent with
$\partial_{\mu}(\rho\sqrt{g}\Sigma^{\mu\nu})=0$ because of
antisymmetry property of the bivector $\Sigma^{\mu\nu}.$
Spherically symmetry property of the space time (\ref{line1})
causes to be just $r$ dependent the string lagrangian density.
Thus we can show that the equation
$\partial_{\mu}(\rho\sqrt{g}\Sigma^{\mu\nu})=0$ reads to the
following solution for the metric equation (\ref{line1}).
\begin{equation}\Sigma^{rt}(r)=\frac{c e^{-A(r)}}{\rho r^2},~~~\Sigma_{rt}(r)=-\frac{c e^{A(r)}}{\rho r^2}
\end{equation}
which by substituting into the lagrangian density (\ref{LNG}) we
obtain
\begin{equation}\label{LNG1}\mathcal{L}_{NG}=\frac{c}{r^2}\end{equation}
 in which $c$ is an integral constant which should be fixed by physical characteristic of the string namely its tension
  $\rho$. To do so we calculate
 the integral equation (\ref{NG}) by using a static gauge \cite{Bar} where the commoving time $\sigma^0$ in the parameter space
 $x^{\mu}(\sigma^0,\sigma^1 )$ of world sheet is equal to the time $t$ in target space as   $\sigma^0=t=t_0=constant$
  and we should assume the static string is along to the radial direction $r$ of world sheet $x^{\mu} (\sigma^0,\sigma^1)$  such that
  \begin{equation}\label{radstring}r=r(\sigma^0,\sigma^1)=F(\sigma^1)
  ,~~~t=\sigma^0,~~~\theta(\sigma^0,\sigma^1)=\varphi(\sigma^0,\sigma^1)=0.\end{equation} Here we
  choose an open string which one edge of the worldsheet
  to be the curve $\sigma^1=0$ and the other edge to be the
curve $\sigma^1=a$ such that $\sigma^1\in[0,a]$ for an open string
with arbitrary shape
  $F(\sigma^1).$
 Anyway if the central object is a
black hole, the string fluid would naturally be attracted/absorbed
by it, and the system would be time-dependent. In order for the
string fluid to be in equilibrium with the black hole, it must
satisfied some specific conditions, like, for example, forming a
disk, and with the strings moving on marginally stable orbits
outside of the event horizon. Even if we assume that the
background metric is spherically symmetric but not static and also
string tension is time dependent there is not a doubt about stable
time independent metric solutions which we considered here because
in Ref.\cite{dharm} the author proved that spherically symmetric
static conditions of a curved spacetime causes to be time
independent the NG string cloud stress tensor and it is a general
solution of the Einstein metric equation. In other words, we have
a "Birkhoff theorem" for the cloud of strings and  so the metric
solution is the general solution for the symmetry under
consideration.
  In this case non-vanishing components of the induced metric (\ref{met3}) reads
   \begin{equation}\label{met4} \mathfrak{g}_{11}=-e^{2A(r)}\bigg(1-\frac{2M(r)}{r}\bigg),~~
   ~~~\mathfrak{g}_{22}=\frac{1}{\big(1-\frac{2M(r)}{r}\big)}\bigg(\frac{dF(\sigma^1)}{d\sigma^1}\bigg)^2,\end{equation}$$
    ~~~\sqrt{\mathfrak{g}}=\sqrt{|det(\mathfrak{g})|}=e^A\frac{dF(\sigma^1)
   }{d\sigma^1}.
 $$ By substituting (\ref{radstring}) and (\ref{met4})
   into (\ref{NG}) we integrate on the worldsheet $\Sigma$ as
   \begin{equation}\label{NG1}I_{\textrm{NG}}=-\int_0^{t}
  \textrm{d}t\int_0^{a}\rho e^A\bigg(\frac{dF(\sigma^1)}{d\sigma^1}\bigg)
   \textrm{d}\sigma^1=-\int_0^t\textrm{d}t\int_0^{r(a)}\rho
   e^A\textrm{d}r.\end{equation}
We should now obtain changed form of the above equation from the
parameter space of worldsheet to the target spacetime
(\ref{line1}). This is done by replacing 4D covariant differential
volume element for the line element (\ref{line1}) given by
   \begin{equation}\label{vol}\textrm{d}v^{[4]}=\sqrt{g}\textrm{d}x^4=4\pi r^2
   e^A\textrm{d}r\textrm{d}t\end{equation} with two dimensional parameter differential surface
    $\textrm{d}\sigma^0\textrm{d}\sigma^1\equiv e^A\textrm{d}t\textrm{d}r$
in the above equation. In this sense we obtain
   \begin{equation}\label{NG2}I_{\textrm{NG}}=-\int\bigg(\frac{\rho}{4\pi r^2}\bigg)\textrm{d}v^{[4]}\end{equation}
which by comparing (\ref{LNG1}) we infer that
\begin{equation}\label{LNG2}\mathcal{L}_{\textrm{NG}}=\frac{\rho}{4\pi r^2},~
~~c=\frac{\rho}{4\pi.}\end{equation}
   By substituting  (\ref{EGB}), (\ref{def}) and (\ref{LNG2}) into the total action functional (\ref{action})
and by integrating angular parts  on the 2-sphere
    $0\leq\theta\leq\pi$, $0\leq\varphi\leq2\pi$
   we obtain \begin{equation}\label{acttot}I=\frac{1}{4G}\int \textrm{d}t\int \textrm{d}r r^2e^{A(r)}\bigg\{
   -\frac{4M^\prime}{r^2}-2\Lambda+12q^2-\frac{\rho}{4\pi r^2}\end{equation}$$
  -\tilde{\alpha}\bigg[\frac{176M^{\prime}q^2}{3r^2}+136q^4+\frac{24M^2}{r^6}-\frac{48MM^\prime}{r^5}\bigg]\bigg\}.$$
 Euler Lagrange equation for $q$ reads
\begin{equation}q\bigg[12-\tilde{\alpha}\bigg(\frac{176M^\prime}{3r^2}+136q^2\bigg)\bigg]=0\end{equation}
which has two different solutions as
\begin{equation}\label{qq}
q_1=0,~~~q_2=\frac{\pm1}{\sqrt{136}}\sqrt{\frac{12}{\tilde{\alpha}}-
\frac{176M^\prime}{3r^2}}.\end{equation}
 Substituting $q_{1,2}$ the equation (\ref{q}) and the action functional (\ref{acttot})
 read to the following forms respectively
\begin{equation}\label{lambda}\lambda^{(1)}_{\textrm{GF}}(r)\sim\frac{1}{r}\end{equation}
\begin{equation}\lambda_{\textrm{GF}}^{(2)}(r)=\int^r\frac{\textrm{d}r^\prime}{r^{\prime2}}
\int^{r^\prime}(r^{\prime\prime})^2q_2(r^{\prime\prime})
e^{A(r^{\prime\prime})}\textrm{d}r^{\prime\prime}
\end{equation} and
\begin{equation}\label{acttot1}I_1=I_2=\frac{1}{4G}\int \textrm{d}t\int \textrm{d}r r^2e^{A(r)}\bigg\{
-\frac{4M^\prime}{r^2}-2\Lambda-\end{equation}$$\frac{\rho}{4\pi
r^2}+\tilde{\alpha}\bigg[-\frac{24M^2}{r^6}+\frac{48MM^\prime}{r^5}\bigg]\bigg\}.$$
    The
   Euler Lagrange equations for the function $A(r)$ and the
   mass distribution function $M(r)$ reduce to the following
   relations respectively.
\begin{equation}\label{N11}
  \frac{4M^\prime}{r^2}=\frac{-\frac{\rho}{4\pi r^2}-2\Lambda-\frac{24\tilde{\alpha}M^2(r)}{r^6}}{1-\frac{12\tilde{\alpha}M(r)}{r^3}}\end{equation}
and
\begin{equation}\label{M11}A^\prime(r)=\frac{\frac{-24\tilde{\alpha}M}{r^4}}{1-\frac{12\tilde{\alpha}M(r)}{r^3}}.\end{equation}
Now we are in position to solve the above nonlinear differential
equations. This is done via numerical approach in this work.
 To do so we need
physical boundary conditions as follows. We know that the mass
distribution function $M(r)$ is related to the matter density
function $\rho_M$ as
 \begin{equation}\label{masin}M(r)=\int_0^r4\pi r^2\rho_{\textrm{M}}(r)\textrm{d}r
 \end{equation}
for which we can write
\begin{equation}\label{masden}\rho_{\textrm{M}}=\frac{M^\prime}{4\pi r^2}.\end{equation}
On the other side for outside region of the
 gravitational compact object (the vacuum zone) with radius $R$ we have
 \begin{equation}\rho_{\textrm{M}}(r)=0,~~~r>R.\end{equation}
While for any arbitrary form of the density function
$\rho_{\textrm{M}}(r)$ the mass integral equation (\ref{masin})
shows the following boundary condition
\begin{equation}\label{mzero}M(0)=0\end{equation}
   To study dynamics of the compact object it is useful we make dimensionless the equation (\ref{N11}) as
   follows.
  \begin{equation}\label{mdot}\dot{m}(y)=
 \frac{-\frac{\rho}{16\pi}-\frac{\Lambda b^2}{2}y^2-\frac{m^2}{y^4}}{1-\frac{2m}{y^3}}\end{equation} where
dot $\dot{}$ is differentiation with respect to $y$ and
  we defined
  \begin{equation}\label{deff}m(y)=\frac{M(r)}{b},~~~y=\frac{r}{b},~~~b=\sqrt{6\tilde{\alpha}}.
 \end{equation}
In this case dimensionless form of the mass density (\ref{masden})
and the equation (\ref{M11}) reduce to
   the following forms respectively. \begin{equation}\label{dens}D_\textrm{M}=8\pi b^2\rho_{\textrm{M}}(y)=
   \frac{\dot{m}}{y^2}=\frac{-\frac{\rho}{16\pi y^2}-\frac{\Lambda b^2}{2}-\frac{m^2}{y^6}}{1-\frac{2m}{y^3}}\end{equation}
    and
    \begin{equation}\label{adot}\dot{A}(y)=\frac{-\frac{4m}{y^4}}{1-\frac{2m}{y^3}}.\end{equation}
   We end this section of the paper by giving the radius of the gravitational compact object and corresponding Horizons.
   Position of radius is obtained by solving $\dot{m}=0$
  such that
   \begin{equation}y_R=\frac{\pm1}{4}\sqrt{\frac{-\rho\pm\sqrt{\rho^2-128\pi^2\Lambda b^2}}{\pi\Lambda b^2}}\end{equation} and
   its horizons are obtained by substituting the horizon hypersurface $y=2m(y)$ into the equation (\ref{mdot})
   as
  \begin{equation}\label{Hor}{y_\textrm{H}}_{\pm}=\frac{\pm1}{\sqrt{1+\frac{\rho}{8\pi}\mp\sqrt{\big(1+
  \frac{\rho}{8\pi}\big)^2+2b^2\Lambda}}}.\end{equation}
By substituting the definition
\begin{equation}\ell=\frac{y_\textrm{H}}{y_\textrm{R}}\end{equation} into the above relations we can obtain
\begin{equation}\label{ell}\frac{\rho}{4\pi}=\frac{\ell^4-2y_\textrm{H}^2+1}{y_\textrm{H}^2(1-\ell^2)},~
~~2b^2\Lambda=\frac{\ell^2[\ell^2-2y_\textrm{H}^2+1]}{y_\textrm{H}^4(\ell^2-1)}\end{equation}
where we must be choose $0<\ell<1$ for a star solution and
$\ell>1$ for a black hole solution. Applying the positivity
condition of the string tension $\rho\geq0$ on the above relations
we obtain
\begin{equation}Star:~~~~~ \ell<1,~~~2y_\textrm{H}^2<1+\frac{1}{\ell^4}\end{equation}
$$Black~ hole:~~~\ell>1,~~~2y_\textrm{H}^2>1+\frac{1}{\ell^4}.$$ It is easy to see that for dS sector $\Lambda>0$
the positive sign in the above horizon positions are valid but in
the case of AdS $\Lambda<0$ both of the positive and negative
signs may be valid and reduce to two different real values. In the
latter case we call smaller horizon to be black hole horizon and
 the larger one to be the modified cosmological horizon. Without
 the cosmological parameter $\Lambda=0$ the above equation gives
 us a curved space time with one horizon such that
 \begin{equation}\label{yH0}y_\textrm{H}=\sqrt{\frac{4\pi}{8\pi+\rho}},~~~\Lambda=0,~~~y_\textrm{R}=\frac{y_\textrm{H}}{
 \sqrt{2y_\textrm{H}^2-1}}.\end{equation}
    We now apply to solve the above dynamical equations for
    different situations as follows.
    \section{Solutions with $\rho=0,~\Lambda(>,<,=)0$}
  The equation (\ref{mdot}) has an analytic exact solution in absence of the string tension  as
  \begin{equation}\label{zeta}m(y)=\zeta y^3,~~~\zeta=\frac{3\pm\sqrt{9+10b^2\Lambda}}{10},~~~\rho=0\end{equation} for which
  the equation (\ref{adot}) reads
  \begin{equation}A(y)=\bigg(\frac{4\zeta}{2\zeta-1}\bigg)\ln(y)\end{equation}
    with corresponding metric components
  \begin{equation}g^{rr}(y)=1-2\zeta y^2,~~~g_{\textrm{tt}}(y)=-y^{\frac{8\zeta}{2\zeta-1}}(1-2\zeta y^2).\end{equation}
  In this case for $2\zeta>1$ we have one apparent horizon with
  position $y_{\textrm{AH}}=\frac{1}{\sqrt{2\zeta}}$ and two event horizons with positions $y_{\textrm{EH}}=0$
  and $y_{\textrm{EH}}=y_{\textrm{AH}}.$
   For $0<2\zeta<1$ position of the event horizon $y_{\textrm{EH}}$ moves to infinity
   $y_{\textrm{EH}}\to\infty$ and (\ref{ell}) reads  \begin{equation}\ell=\bigg(\frac{1}{\zeta}-1\bigg)^\frac{1}{4}.\end{equation}
   This shows that for $\zeta>\frac{1}{2}$ we have a black hole in which $\ell<1$ and so $y_\textrm{R}<y_\textrm{H}$ while for
    $0<\zeta<\frac{1}{2}$
   we have $y_\textrm{R}>y_\textrm{H}$ and so the above metric is not a black hole but may to be a regular star.
  The definition of $\zeta$ in the
relation (\ref{zeta}) shows that we must be choose
\begin{equation}\Lambda\geq\frac{-3}{20\tilde{\alpha}}.
  \end{equation} This can be understood as the unknown cosmological constant in general theory of
  relativity originates in fact from the GB coupling parameter and so can be understandable.
  By according to the relationship $\tilde{\alpha}=(D-4)\alpha_{\textrm{GB}}$ and positivity condition of
  the GB coupling constant $\alpha_{\textrm{GB}}>0$ we infer that the above relationship  corresponds to a AdS (dS)
   space when $\tilde{\alpha}>0(<0)$
   and so $D>4(<4).$
  This means that when we regularize the GB action functional it should be done by limiting from higher space time dimensions to
$4D$ and vice versa for
  dS space. For the above analytic metric solution it is easy to show that the Hawking temperature
  $\bar{T}=bT=-\frac{1}{2\pi}\frac{\textrm{d}g_{\textrm{tt}}(y)}{\textrm{d}y}_{|y_\textrm{H}}$
  reads \begin{equation}\label{temzero}\bar{T}=2(2\zeta)^{\frac{1}{2}\big(
  \frac{1+6\zeta}{1-2\zeta}\big)}.\end{equation}
   We know that at the extended phase space, the pressure of AdS space is defined by
   $P=-\frac{\Lambda}{8\pi}$ for negative cosmological constant $\Lambda<0$
   and $P=\frac{\Lambda}{8\pi}$ for dS sector with $\Lambda>0$.By regarding these definitions the parameter of $\zeta$
given by (\ref{zeta})
reads\begin{equation}\label{perzero}\bar{P}=b^2P=\frac{\pm[(10\zeta-3)^2-9]}{80\pi}\end{equation}
where $+(-)$ corresponds to dS(AdS) sector. For pressureless space
$\Lambda=0$ we have $\zeta=\{0,0.6\}$ for which the corresponding
temperature will be $T(0)=0$ and $T(0.6)=0.1422.$ Metric field
solution is flat Minkowski for $\zeta=0$ but not for $\zeta=0.6.$
Figure 1-a shows event and apparent horizons of the space time in
the latter case which their positions are crossing points with the
horizontal axes. Event horizon is obtained by solving
$g_{\textrm{tt}}(y)=0$ and apparent horizon is obtained with
$g^{\textrm{rr}}(y)=0$ for spherically symmetric state space
times. Pressure-temperature phase diagram for the equations
(\ref{temzero}) and (\ref{perzero}) are plotted in the figure
1-b,1-c,1-d. These diagrams show a dS/AdS phase transition with
coexistence state (butterfly tail form in figures) between them at
the crossing point in the P-T diagrams.
\section{Initial conditions with $\rho>0,~\Lambda(>,<,=)0$}
For the case $\rho\neq0$ the equation (\ref{mdot}) has not
analytic solution and it should be solved via numerical methods.
To do so  we apply Runge Kutta method for which we should assume
some physical initial conditions for $m(y)$, $\rho$ and $\Lambda$
and domain of $y$. By looking at the equation (\ref{mzero}) one
can infer that a suitable initial condition for the mass parameter
is
\begin{equation}\label{bound1}m(0)=0\end{equation} while we are still free to choose various
values for $\Lambda,$ $\rho$ and regimes of the variable $y.$ To
determine that what is suitable regimes for these parameters? we
call equation of state of the system by calculating the
corresponding Hawking temperature as follows. We know that the
Hawking temperature of a black hole space time is determined by
value of surface gravity on its exterior horizon such that
\begin{equation}\label{Haw}T=\frac{-1}{4\pi}\frac{\textrm{d}
g_{\textrm{tt}}(r)}{\textrm{d}r}_{\big|r=2M(r)}=\frac{e^{2A(r)}}{2\pi}\bigg[A^\prime(r)\bigg(1-\frac{2M(r)}{r}\bigg)+
\frac{M(r)}{r^2}-\frac{M^\prime(r)}{r}\bigg]_{\big|_{\textrm{r=2M(r)}}}.
\end{equation} For  AdS (dS) background with pressure $P=\frac{-\Lambda}{8\pi}>0~ (P=\frac{\Lambda}{8\pi}>0)$ we substitute (\ref{N11}) and
(\ref{M11}) and definition (\ref{deff}) into the Hawking
temperature (\ref{Haw}) to obtain the 4DGB black hole equation of
state such that
\begin{equation}\label{es1}AdS:~~~\bar{T}=\bar{P}\bar{v}-\bar{v}f(\bar{v}),~~~~\bar{v}=\frac{v}{b}=\frac{2y^3e^{2A(y)}}{1-y^2},~~~0\leq
y\leq1\end{equation}
and\begin{equation}\label{es2}dS:~~~\bar{T}=\bar{P}\bar{v}+\bar{v}f(\bar{v}),~~~~\bar{v}=\frac{v}{b}=\frac{2y^3e^{2A(y)}}{y^2-1},~~~y\geq1\end{equation}
where $\bar{v}$ is dimensionless specific volume and
\begin{equation}f[\bar{v}(y)]=\frac{1}{4\pi
y^2}\bigg[\frac{1}{2}+\frac{\rho}{16\pi}-\frac{1}{4y^2}
\bigg].\end{equation} By looking at the above equation of state
one can infer that at large scales $y\to\infty$ the function
$f(\bar{v})$ vanishes and so the above equation of state reaches
to the well known ideal gas equation of state.  Now we are in
position to obtain critical point in T-v extended phase space by
calculating equation of the critical points $\frac{\partial
T}{\partial v}{\big|_{P}}=0$ and $\frac{\partial^2 T}{\partial
v^2}{\big|_{P}}=0$ which reduce to the following conditions.
\begin{equation}\frac{\partial \bar{T}}{\partial y}{\big|_{P}}=0,~~~\frac{\partial^2 \bar{T}}{\partial y^2}{\big|_{P}}=0\end{equation}
The above equations give us a parametric critical point such that
for AdS we have
\begin{equation}\label{P22}AdS:~~~
\bar{P}_\textrm{c}(y_\textrm{c})=\frac{-3(y_\textrm{c}^4+2y_\textrm{c}^2+5)}{16\pi
y_\textrm{c}^4(y_\textrm{c}^4+10y_\textrm{c}^2-35)},\end{equation}
$$
\bar{v}_\textrm{c}=\frac{2y_\textrm{c}^3e^{2A(y_\textrm{c})}}{1-y_\textrm{c}^2},~~~\bar{T}_\textrm{c}=\frac{(y_\textrm{c}^2-1)e^{2A(y_\textrm{c})}}{\pi
y_\textrm{c}(y_\textrm{c}^4+10y_\textrm{c}^2-35)}$$ and for dS
\begin{equation}\label{P2}dS:~~~\bar{P}_\textrm{c}(y_\textrm{c})=\frac{3(y_\textrm{c}^4+2y_\textrm{c}^2+5)}{16\pi
y_\textrm{c}^4(y_\textrm{c}^4+10y_\textrm{c}^2-35)}\end{equation}$$
\bar{v}_\textrm{c}=\frac{2y_\textrm{c}^3e^{2A(y_\textrm{c})}}{y_\textrm{c}^2-1},~~~\bar{T}_\textrm{c}=\frac{(y_\textrm{c}^4-14y_\textrm{c}^2-11)e^{2A(y_\textrm{c})}}{4\pi
y_\textrm{c}(y_\textrm{c}^2-1)(y_\textrm{c}^4+10y_\textrm{c}^2-35)}$$
where for both sectors dS and AdS the string tension satisfies to
the following relation.
\begin{equation}\rho(y_\textrm{c})=\frac{-8\pi(y_\textrm{c}^6+7y_\textrm{c}^4-29y_\textrm{c}^2+21)}{y_\textrm{c}^2(y_\textrm{c}^4+10y_\textrm{c}^2-35)},\end{equation}
for arbitrary parameter $y_\textrm{c}.$ By looking at the figure
2-a one can obtain that $\rho=0$ at $y_\textrm{c}=\pm1$ and
$y_\textrm{c}=\pm1.4432.$ Furthermore $\rho>0$ for
$0<y_\textrm{c}<1$ and $1.4432<y_\textrm{c}<1.6448.$ These regions
are positioned in the AdS and dS regimes respectively. and satisfy
positivity condition on the string tension. Thus to set suitable
numeric critical points form the above parametric form consisted
with dS and AdS regions we choose the following ansatz.
\begin{equation}\label{AdScon}AdS:~y_\textrm{c}=\frac{1}{2},~\bar{T_\textrm{c}}=0.015e^{2A(0.5)},~\bar{P}_\textrm{c}=0.16,~\rho=44.02,~\bar{v}_\textrm{c}=0.33e^{2A(0.5)}\end{equation} and
\begin{equation}\label{dScon}dS:~y_\textrm{c}=\frac{3}{2},~\bar{T_\textrm{c}}=0.21e^{2A(1.5)},~\bar{P}_\textrm{c}=-0.023,~\rho=3.87,~\bar{v}_\textrm{c}=5.4e^{2A(1.5)}\end{equation} where numeric
values for $A(0.5)$ and $A(1.5)$ can be obtained by solving the
equations (\ref{mdot}) and (\ref{adot}). By substituting the above
critical points into the equations (\ref{ell}) we obtain two class
of solutions which one of them is star $\ell<1$ and the other one
is a black hole $\ell>1$ as follows.
\begin{equation}\label{oo}AdS:~~\{y_\textrm{H}\approx0.43,~~ y_\textrm{R}\approx16.60\}_{\textrm{star}},~~
\{y_\textrm{H}\approx20.78,~~
y_\textrm{R}\approx16.60\}_{\textrm{black~hole}}\end{equation}
\begin{equation}\label{pp}dS:~~\{y_\textrm{H}\approx0.66,~~y_\textrm{R}\approx13.09\}_{\textrm{star}},~~
 \{y_\textrm{H}\approx35.50,~~y_\textrm{R}\approx13.09\}_{\textrm{black~hole}}\end{equation}
where we omitted negative roots of the equations (\ref{ell}).
Because solutions in the regions with $y<0$  are positioned in the
analytic continuation of $y$ in the complex algebraic analysis.
Also we substitute $P_\textrm{c}=\frac{\Lambda}{8\pi}$ and
$P_\textrm{c}=-\frac{\Lambda}{8\pi}$ into the above critical
points to obtain
\begin{equation}\label{Lambdacon}b^2\Lambda_{\textrm{AdS}}=-0.006,~~~b^2\Lambda_{\textrm{dS}}=-0.0009.\end{equation}
for the dS and the AdS cases respectively. We should remember that
the black hole choice in (\ref{oo}) and the star choice in
(\ref{pp}) are not physical because for them the thermodynamic
specific volume takes on negative sign. Thus in the next section
we apply to solve (\ref{mdot}) and (\ref{adot}) just for star
choice in the AdS and the black hole choice in the dS sector.
\section{Numerical analysis} In the
section 3 we obtained that the spatial regimes $0\leq y<\infty$
are separated into two subregions such that for dS (AdS) sector we
should use $y>1 (0<y<1)$ by omitting analytic continuation regions
$y<0.$ Also in that section we said that $m(0)=0$ is suitable
initial condition for mass function when we solve dynamical
equations (\ref{mdot}) and (\ref{adot}) as numerically. In the
previous section we obtained consistent initial conditions on
$\rho$ and $\Lambda$ parameters of the system under consideration.
By substituting (\ref{AdScon}),(\ref{dScon}) and (\ref{Lambdacon})
the equation (\ref{mdot}) reduces to the following form for dS and
AdS sectors.
\begin{equation}dS:~~~\dot{m}=\frac{-0.077+0.0005y^2-\frac{m^2}{y^4}}{1-\frac{2m}{y^3}},~~~y>1\end{equation}
$$ AdS:~~~\dot{m}=\frac{-0.88+0.003y^2-\frac{m^2}{y^4}}{1-\frac{2m}{y^3}},~~~0<y<1$$
for which the equation of state become
\begin{equation}\label{eqsdS}dS:~~~\bar{T}=\bar{v}\bigg[\bar{P}+\frac{1}{4\pi
y^2}\bigg(0.58-\frac{1}{4y^2}\bigg)\bigg],~~~\bar{v}=\frac{2y^3e^{2A(y)}}{y^2-1}
\end{equation}
$$AdS:~~~\bar{T}=\bar{v}\bigg[\bar{P}-\frac{1}{4\pi y^2}\bigg(1.38-\frac{1}{4y^2}\bigg)\bigg],~~~\bar{v}=\frac{2y^3e^{2A(y)}}{1-y^2}$$
where $A(y)$ is obtained by solving
\begin{equation}\dot{A}=\frac{-\frac{4m}{y^4}}{1-\frac{2m}{y^3}}.
\end{equation}
Now we are in position to choose valid domains for $y$ to produce
numeric solutions of the fields. By looking at the (\ref{oo}) and
(\ref{pp}) we choose
\begin{equation}AdS:~~~0<y<1,~~~y_H=0.43,~~~y_c=0.5,\end{equation} and \begin{equation}dS:~~~1<y<40,~
~~y_H=35.50,~~y_c=1.5,~~~y_R=13.09\end{equation} We use Runge
Kutta method via the Maple and the Mathematica softwares. To
produce numeric solutions with this method we should have a
suitable form for the dynamical equations as
$\frac{ds}{dx}=f(x,s).$ In fact the equation (\ref{mdot}) has this
form but (\ref{adot}) has not. To resolve this problem we obtain a
consistent differential equation instead of (\ref{adot}). This is
done for $\dot{A}=w(y)$ just by reproducing $m(y)$ from
(\ref{adot}) and calculating its $\dot{m}$  and by substituting
them into the equation (\ref{mdot}) such that
\begin{equation}\label{wdot}\dot{w}=\frac{0.31}{y^3}-\frac{0.002}{x}-\bigg(5y+\frac{0.5}{y}\bigg)w\end{equation}$$
+y(0.23+1.75y^2)w^2-y^2(0.04+0.13y^2)w^3.$$ To generate a suitable
differential equation for $g^{\textrm{rr}}(y)=S(y)$ we substitute
$m=(1-S)y/2$ and its first derivative into the equation
(\ref{mdot}) such that
\begin{equation}\label{sdot}\frac{dS}{dy}=-\frac{4\alpha y^4+(2S+4\beta-2)+(S-1)^2}{y(y^2+S-1)}\end{equation}
where \begin{equation}dS:~~~\alpha=-0.077,~~~\beta=0.0005
\end{equation} and
\begin{equation}AdS:~~~\alpha=-0.88,~~~~\beta=0.003.\end{equation}By holding the equations (\ref{mdot}), (\ref{wdot}) and (\ref{sdot}), the
Maple software takes out the best fitting numeric solutions for
$m(y), g^{\textrm{rr}}(y)$ and $\dot{A}(y)$ given in the figures
3-a,c,d for dS and 4-a,c,e for AdS sectors. Several points on the
curves generated by the machine are listed in the tables 1 and 2
and we used them to determine numerical values of the fields
$A(y)$ and $g_{\textrm{tt}}(y)$ via Mathematica software.  By
looking at these diagrams one can see that the metric fields in
both cases of dS and AdS have a crossing point with the horizontal
axes which means that they are locations of the black hole
horizon.One can infer the most important result from the volume
pressure phase diagrams at constant temperatures that: by looking
at the figure 3-f we understand that a dS 4DGB black hole
participates in a large to small black hole phase transition for
temperatures less that the critical one. For
$\bar{T}\geq\bar{T}_c$ this black hole at maximal pressure is in a
disequilibrium state and it reaches to a vacuum AdS finally. At
all in both cases of dS and AdS spaces a 4DGB black hole
surrounded with a cloud of string take on two phases which may to
be in a coexistence state when they have small scales but not in
the large scales. In about the AdS 4DGB black hole in presence of
the string tension the figure 4-f shows that this black hole at
$\bar{T}<\bar{T}_c$ is unstable thermodynamically at the maximum
pressure and it can be participated in the Hawking-Page phase
transition where the black hole evaporates to a vacuum dS finally.
For the cases $\bar{T}\geq\bar{T}_c$ this black hole does not take
on a phase transition.
\section{Conclusion}
In this  work we choose EGB gravity model \cite{AL} which is
consistent in 4D curved spacetimes and solved metric equations for
spherically symmetric static black hole line element with and
without the cosmological constant and the Numbo Goto string
tension. In absence of the string tension we obtained analytic
solution for the metric fields but with string tension we used
Runge Kutta method to obtain numeric solutions of the fields. By
studying thermodynamic of these  black holes we infer that for
small scales they behave as two fluid systems which for
temperatures less than the critical one a dS black hole
participates in the large to small black hole phase transition
while AdS one reaches to the Hawking-Page phase transition. As
extension of this work we like to study in our next work
possibility of the Joule-Thomson  expansion of this system and
other thermodynamic behavior of the obtained metric solution.
\begin{center} Table 1. Numerical solutions for dS pressure
\end{center}
\begin{center}
\begin{tabular}{|c|c|c|c|c|c|}
\hline
$y$ & $m$ &$10^{7}\times\dot{A}$& $A$& $g^{rr}$&$g_{tt}$\\
\hline
35& 16.984& 0& 35& 0.0294857& $-7.41695\times10^{28}$\\
40&  20.49& $-1.2216$& 34.99999635& -0.0245& $6.16278\times10^{28}$\\
45&  25.07& $-1.5739$&  34.99999704& -0.114222& $2.87317\times10^{29}$\\
50&  30.24& $-1.60278$&  34.99999780& -0.2096& $5.27234\times10^{29}$\\
55& 36.52& $-1.512854$& 34.99999860& -0.328& $8.25062\times10^{29}$\\
60& 45.03& $-1.38500$&   34.99999939& -0.501& $1.26023\times10^{30}$\\
\hline
\end{tabular}
\end{center}
\begin{center} Table 2. Numerical solutions with
AdS pressure
\end{center}
\begin{center}
\begin{tabular}{|c|c|c|c|c|c|}
\hline
$y$ & $m$ &$\dot{A}$& $A$& $g^{rr}$&$g_{tt}$\\
\hline
0.42& 0.21& 0& 5.78182& 0.& 0.\\
0.54& 0.28& 4.2&  5.15252& -0.037037& 164.706\\
0.62& 0.38& 4.878&  4.69953& -0.225806& 3896.85\\
0.768& 0.506& 5.194&  4.71439& -0.317708& 10315.2\\
0.88& 0.68& 5.52552&  4.55532& -0.545455& 34368.8\\
0.998& 0.897& 5.6506&  4.49406& -0.797595& 64540.4\\
\hline
\end{tabular}
\end{center}
\begin{figure}[ht]
\centering  \subfigure[{}]{\label{101}
\includegraphics[width=0.4\textwidth]{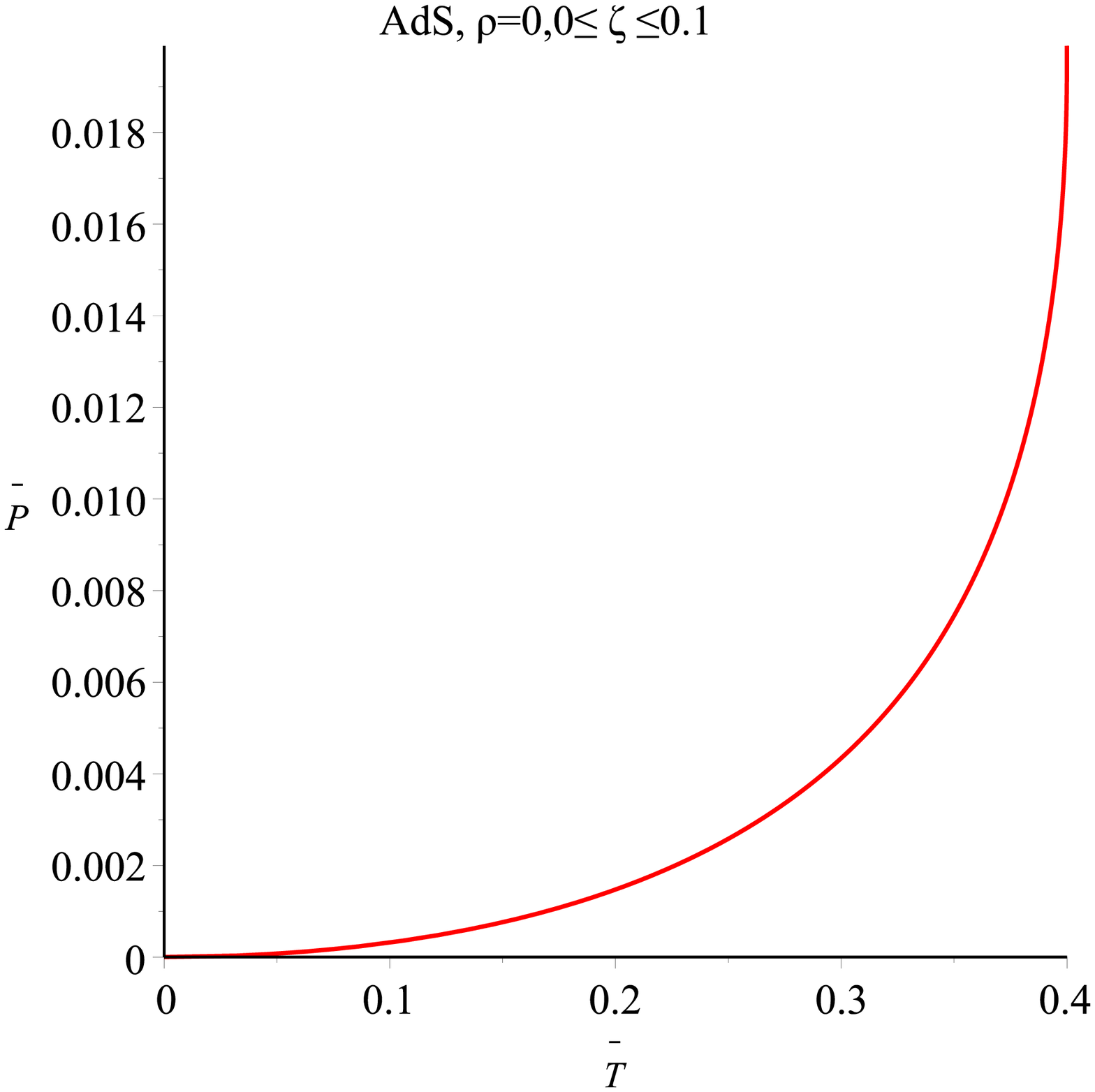}}
\hspace{3mm}\subfigure[{}]{\label{2341}
\includegraphics[width=0.4\textwidth]{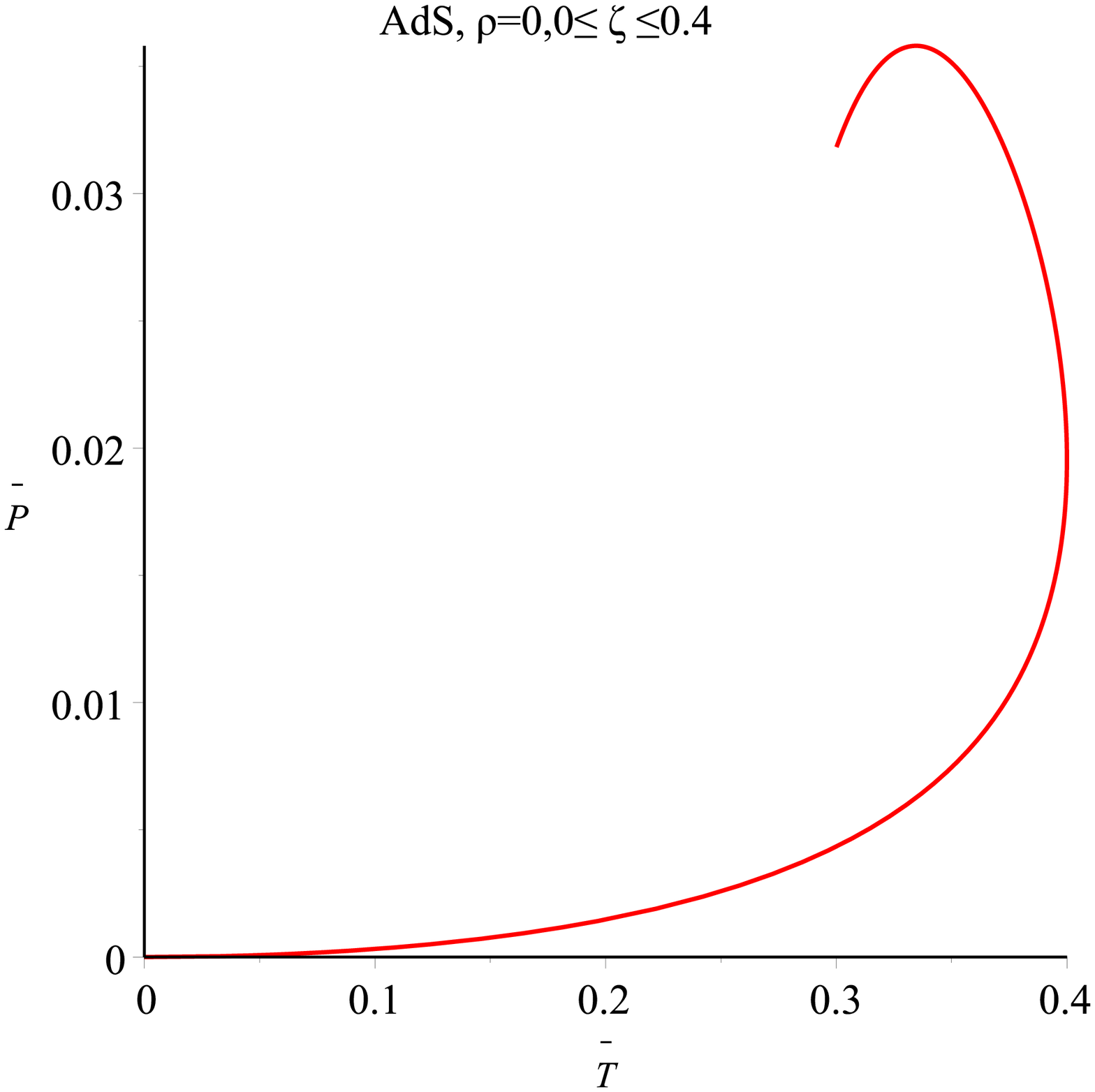}}
\hspace{3mm} \subfigure[{}]{\label{s1}
\includegraphics[width=0.4\textwidth]{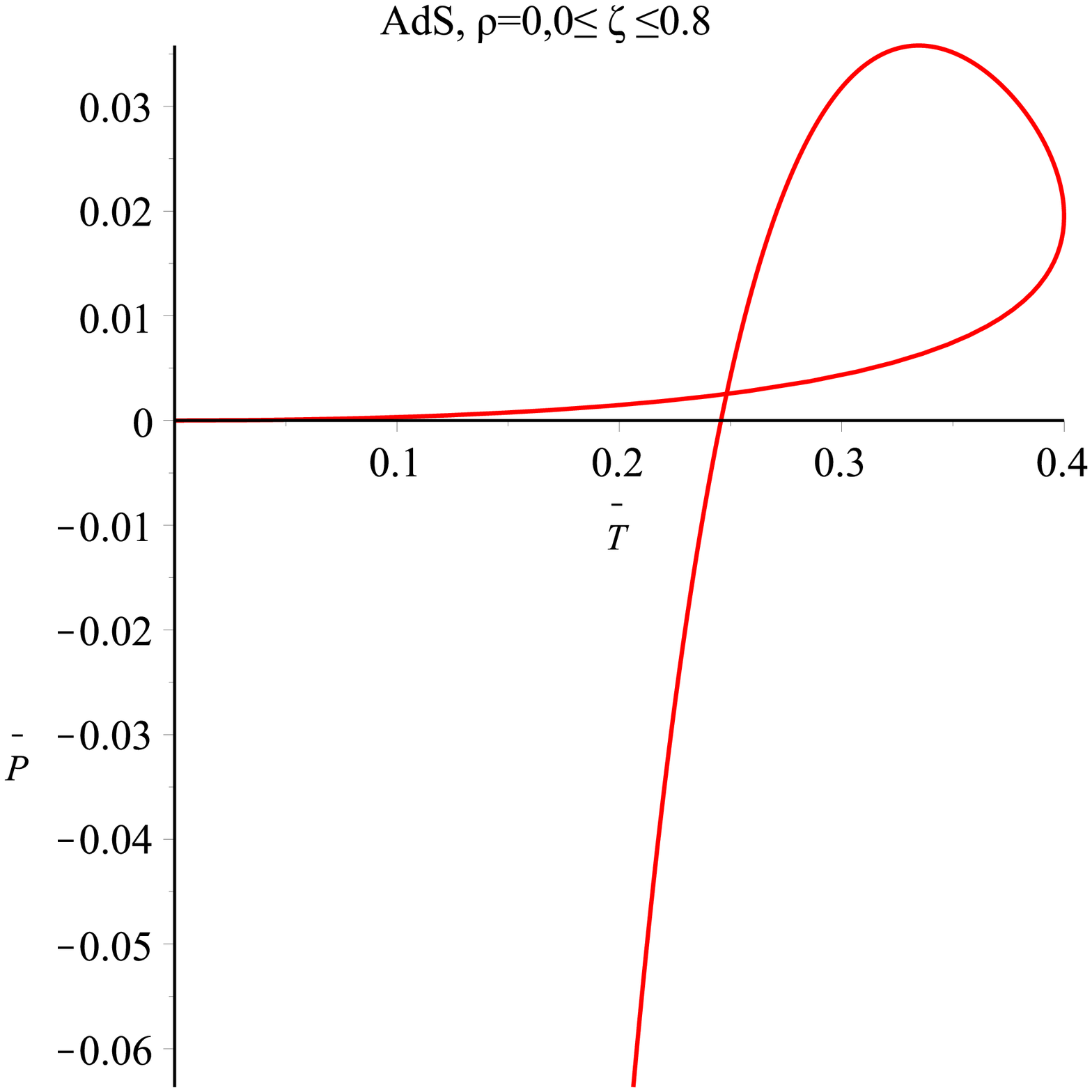}}
\hspace{3mm} \subfigure[{}]{\label{a1}
\includegraphics[width=0.4\textwidth]{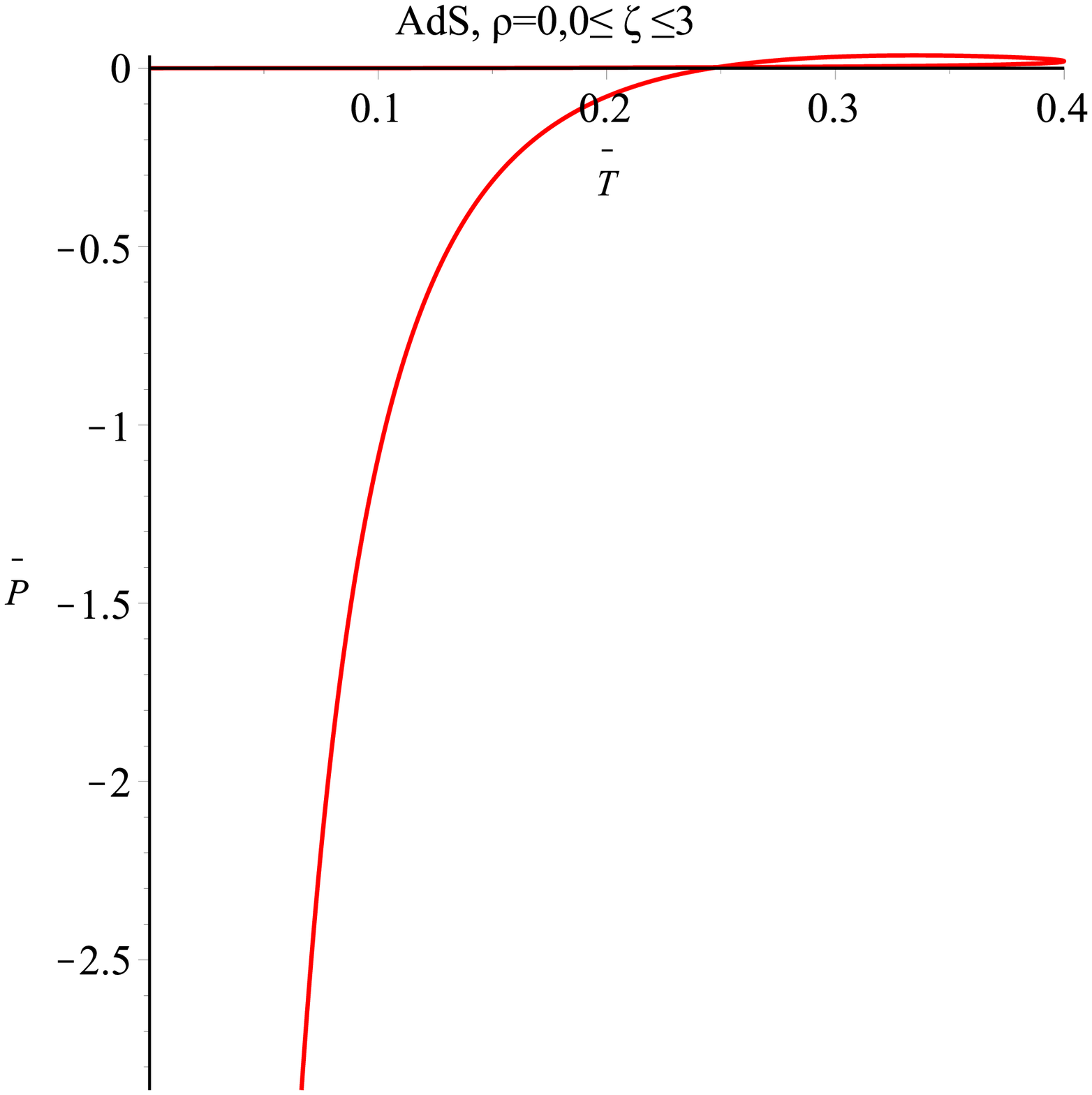}}
\hspace{3mm}  \caption{ P-T diagrams for $\rho=0$ with AdS
background. For dS sector diagrams are similar with these curves
except where the pressures should be inverted as $\bar{P}\to
-\bar{P}$ }
\end{figure}
\begin{figure}[ht]
\centering  \subfigure[{}]{\label{22}
\includegraphics[width=0.4\textwidth]{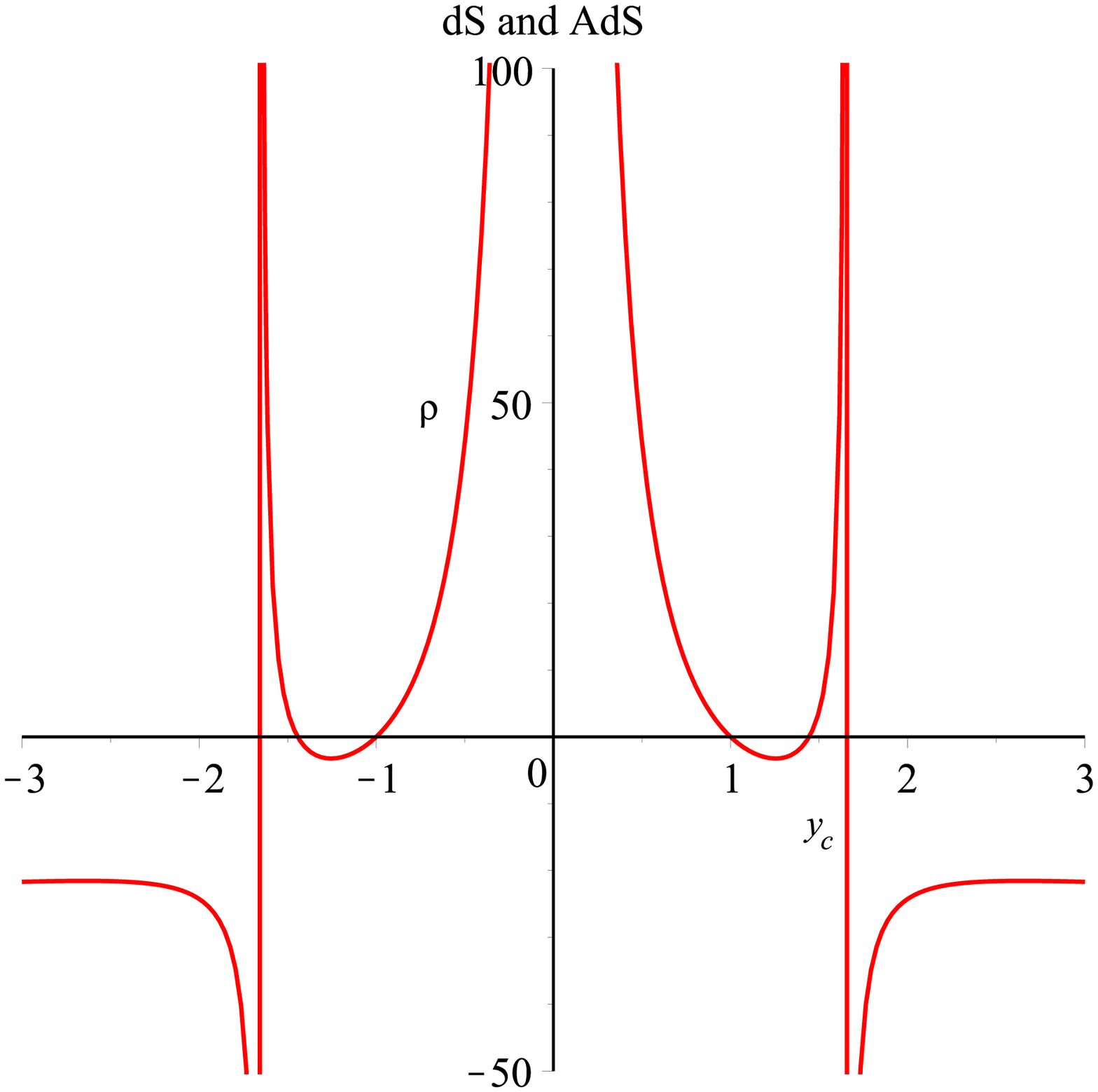}}
\hspace{3mm} \subfigure[{}]{\label{32}
\includegraphics[width=0.4\textwidth]{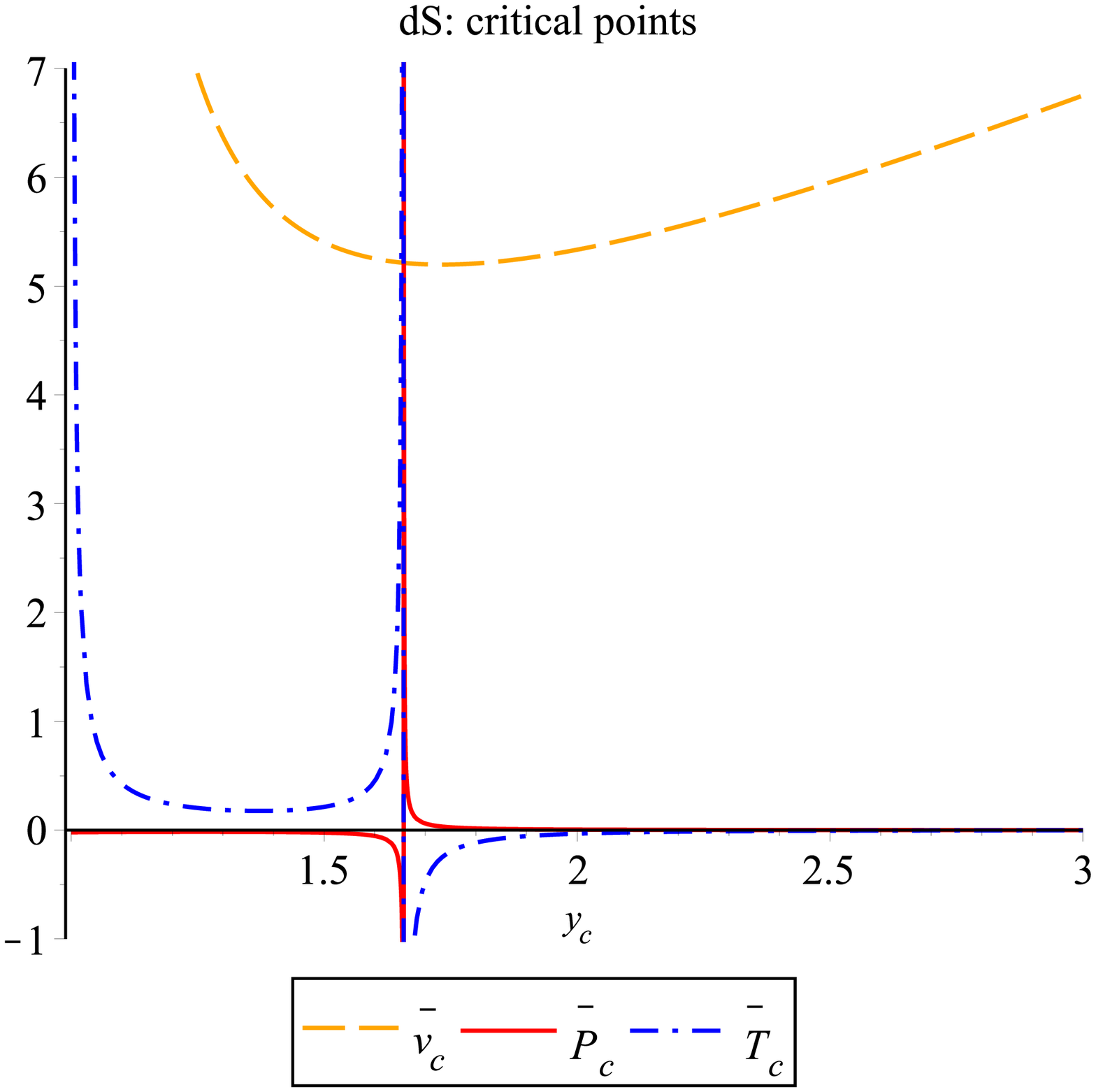}}
\hspace{3mm} \subfigure[{}]{\label{12}
\includegraphics[width=0.4\textwidth]{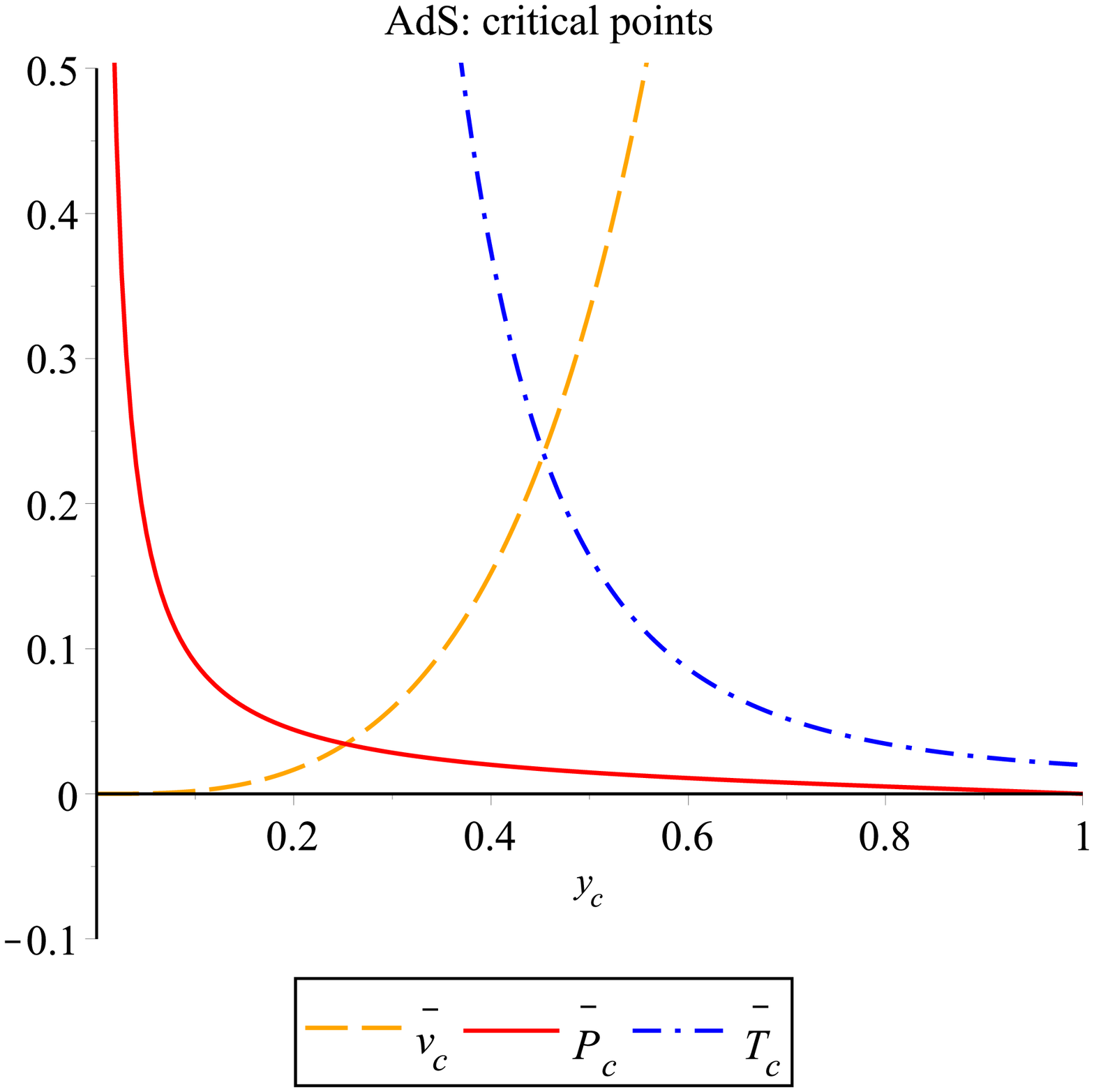}}
\hspace{3mm}  \caption{ Numeric values of the critical points.}
\end{figure}
\begin{figure}[ht]
\centering \hspace{3mm}\subfigure[{}]{\label{3}
\includegraphics[width=0.3\textwidth]{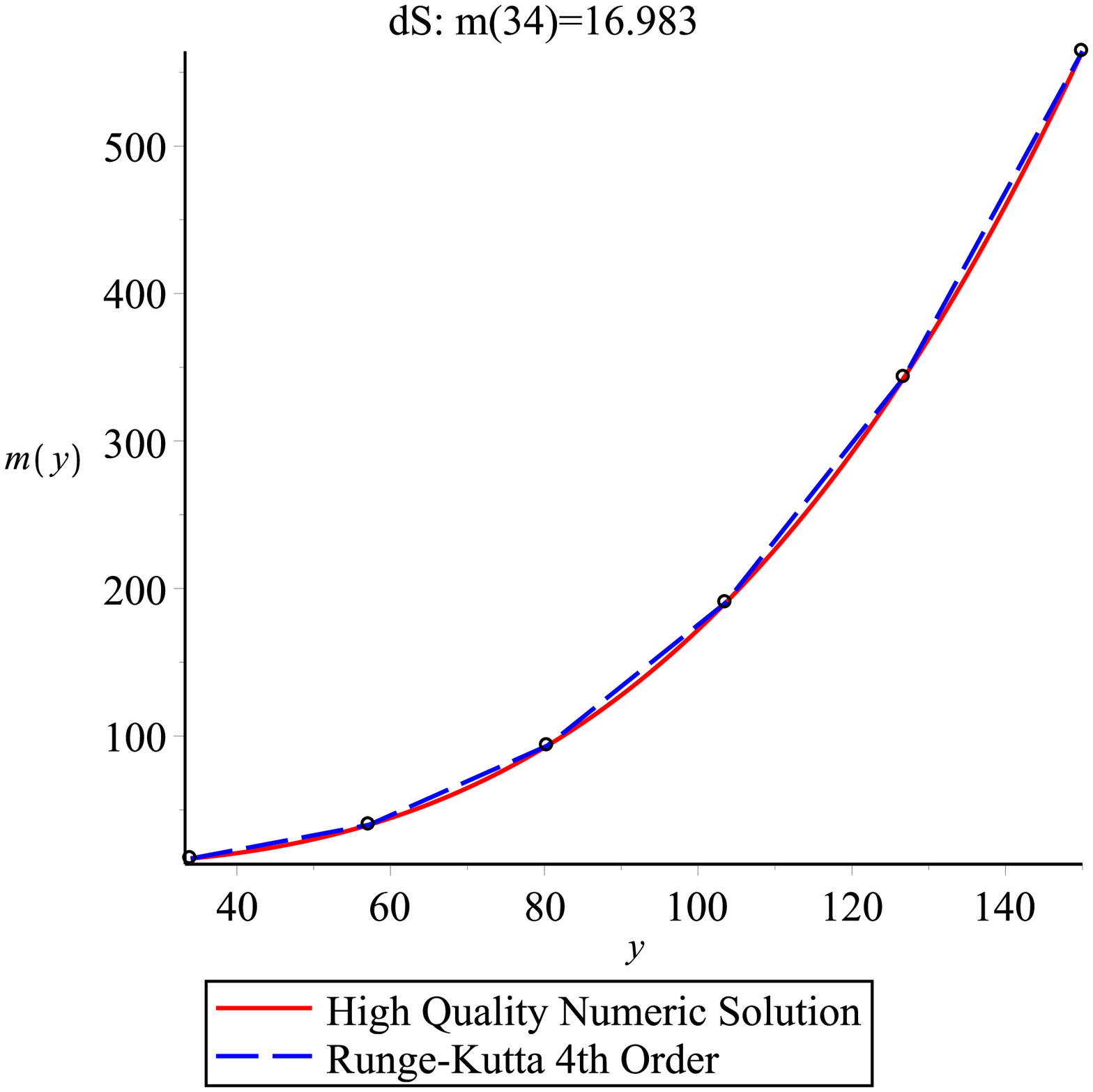}}
\hspace{3mm}\subfigure[{}]{\label{122}
\includegraphics[width=0.3\textwidth]{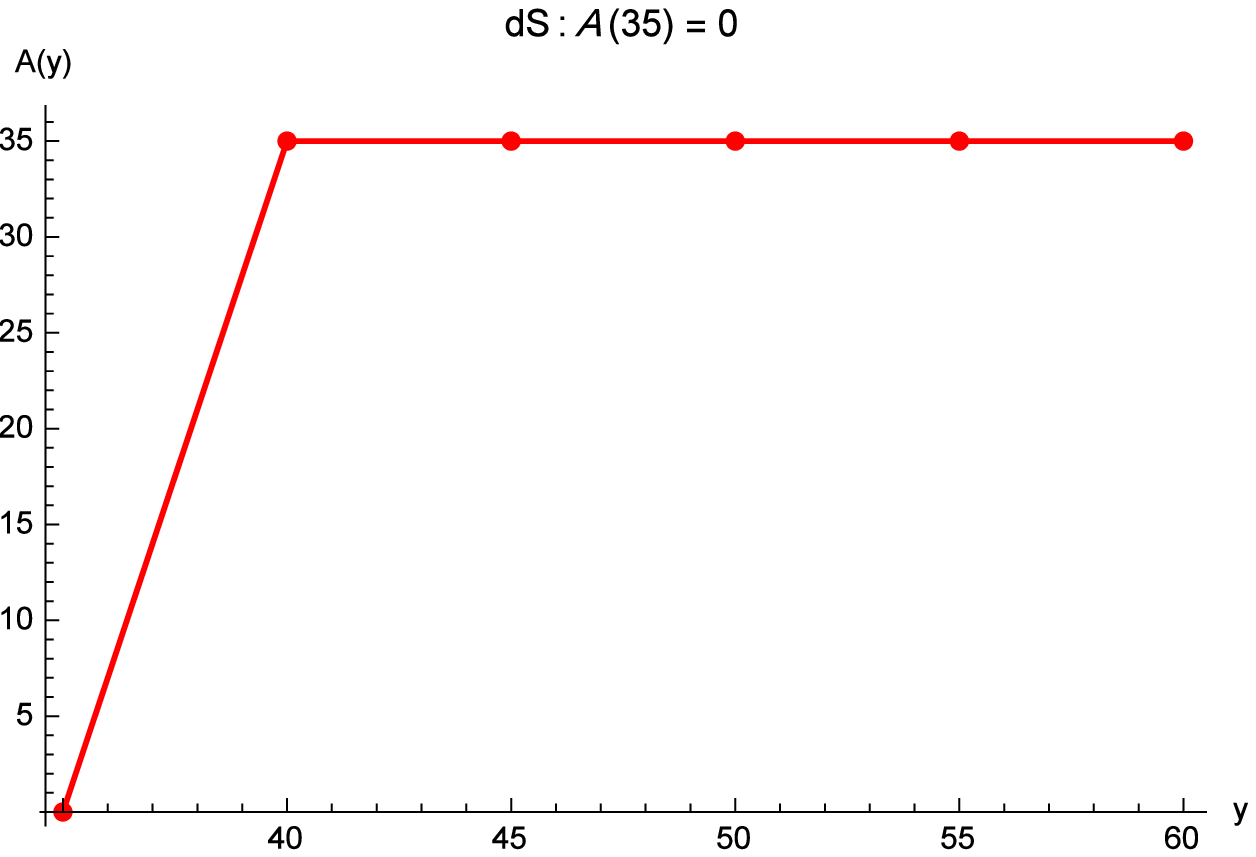}}
\hspace{3mm}\subfigure[{}]{\label{1222}
\includegraphics[width=0.3\textwidth]{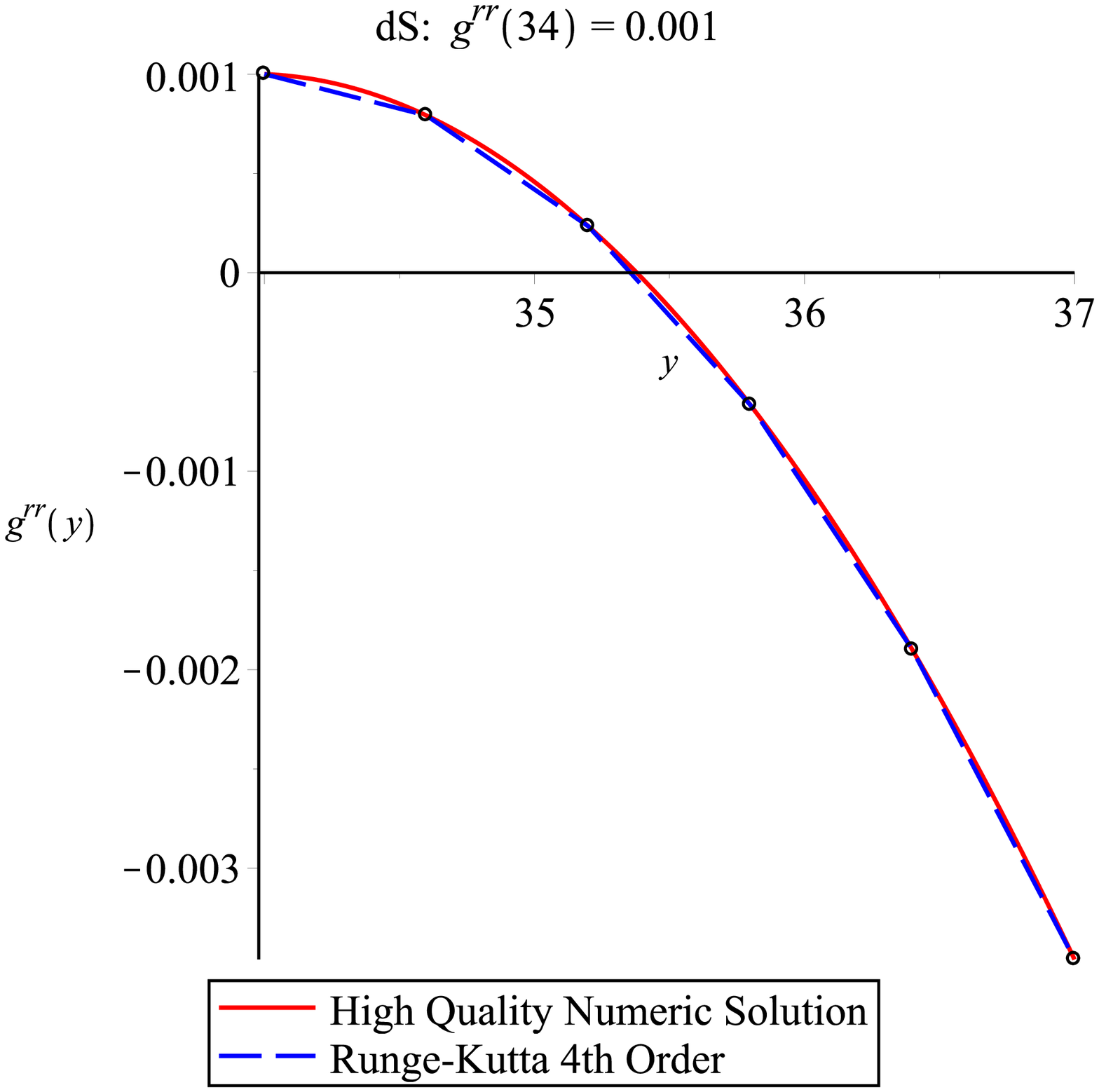}}
\hspace{3mm}\subfigure[{}]{\label{231}
\includegraphics[width=0.3\textwidth]{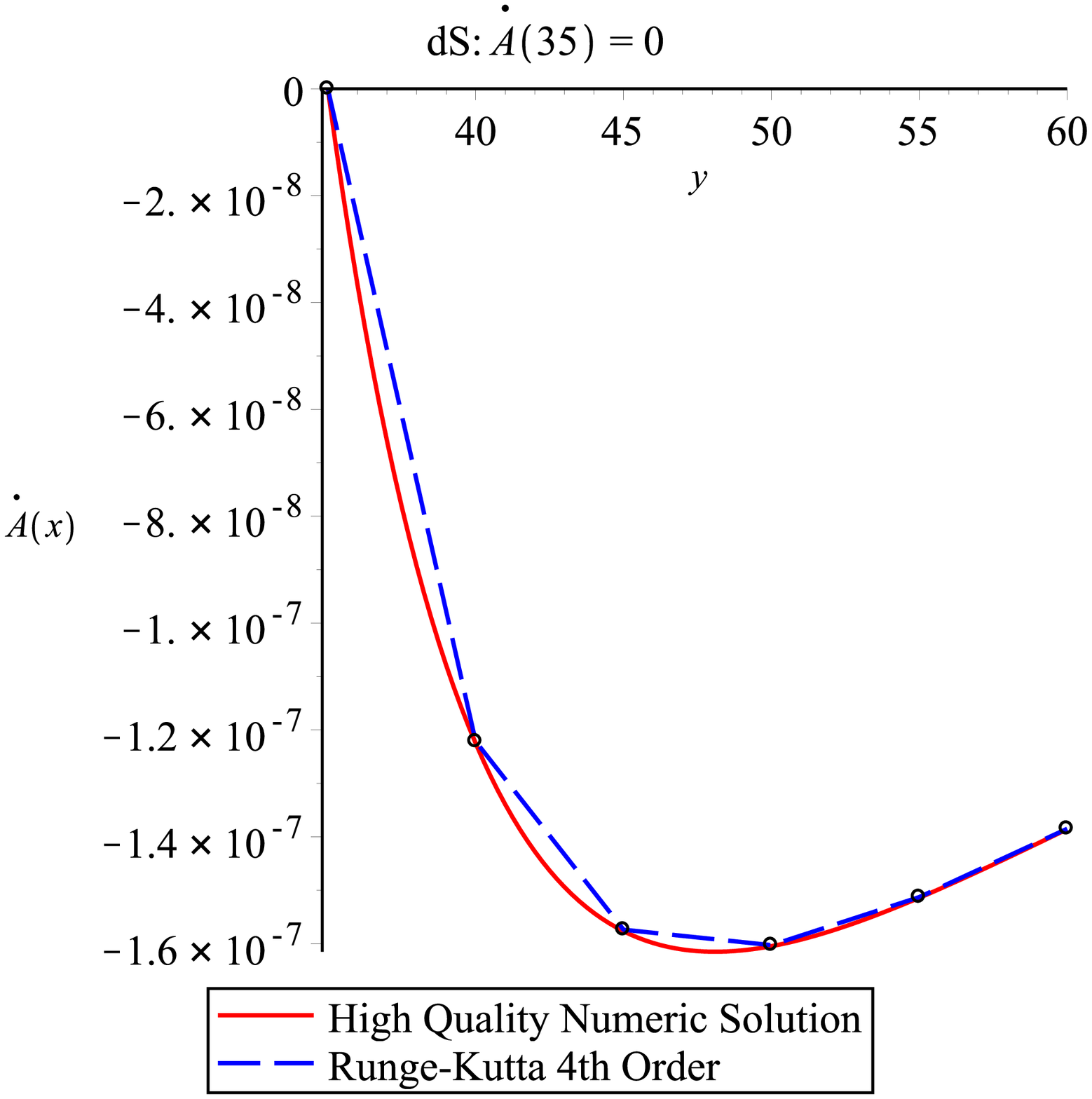}}
\hspace{3mm}\subfigure[{}]{\label{321}
\includegraphics[width=0.3\textwidth]{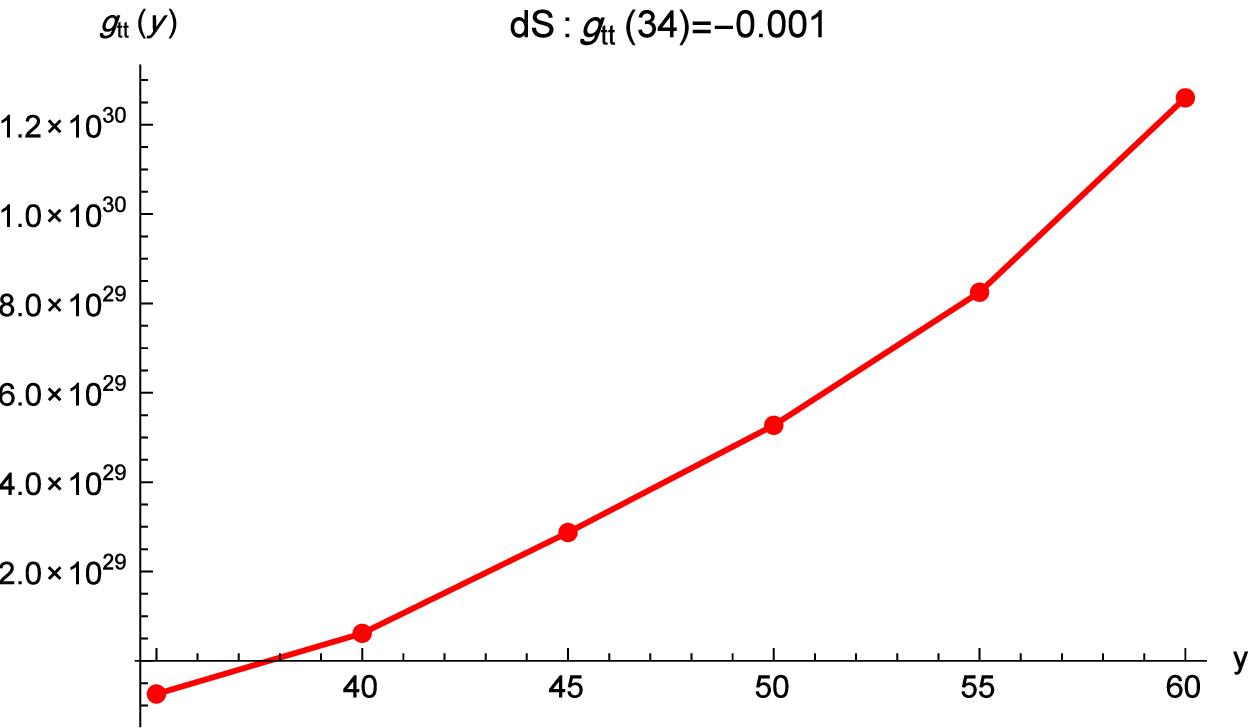}}
\hspace{3mm}\subfigure[{}]{\label{121}
\includegraphics[width=0.3\textwidth]{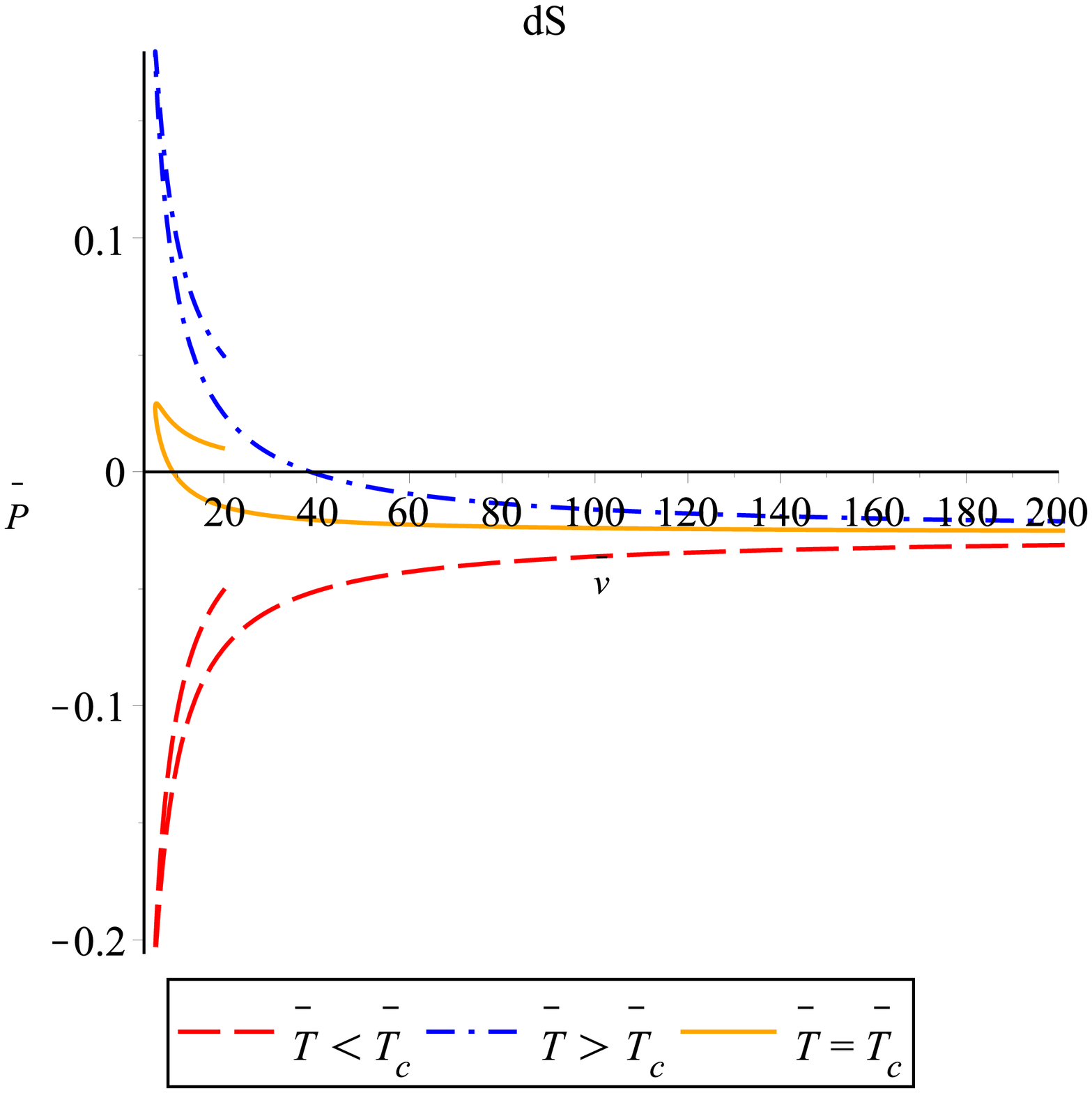}}
\hspace{3mm} \caption{ Diagrams of the numeric solutions of the
fields with the dS background space. Initial values to produce
numerical solutions are shown at top of each diagram. }
\end{figure}
\begin{figure}[ht]
\centering \hspace{3mm}\subfigure[{}]{\label{4}
\includegraphics[width=0.3\textwidth]{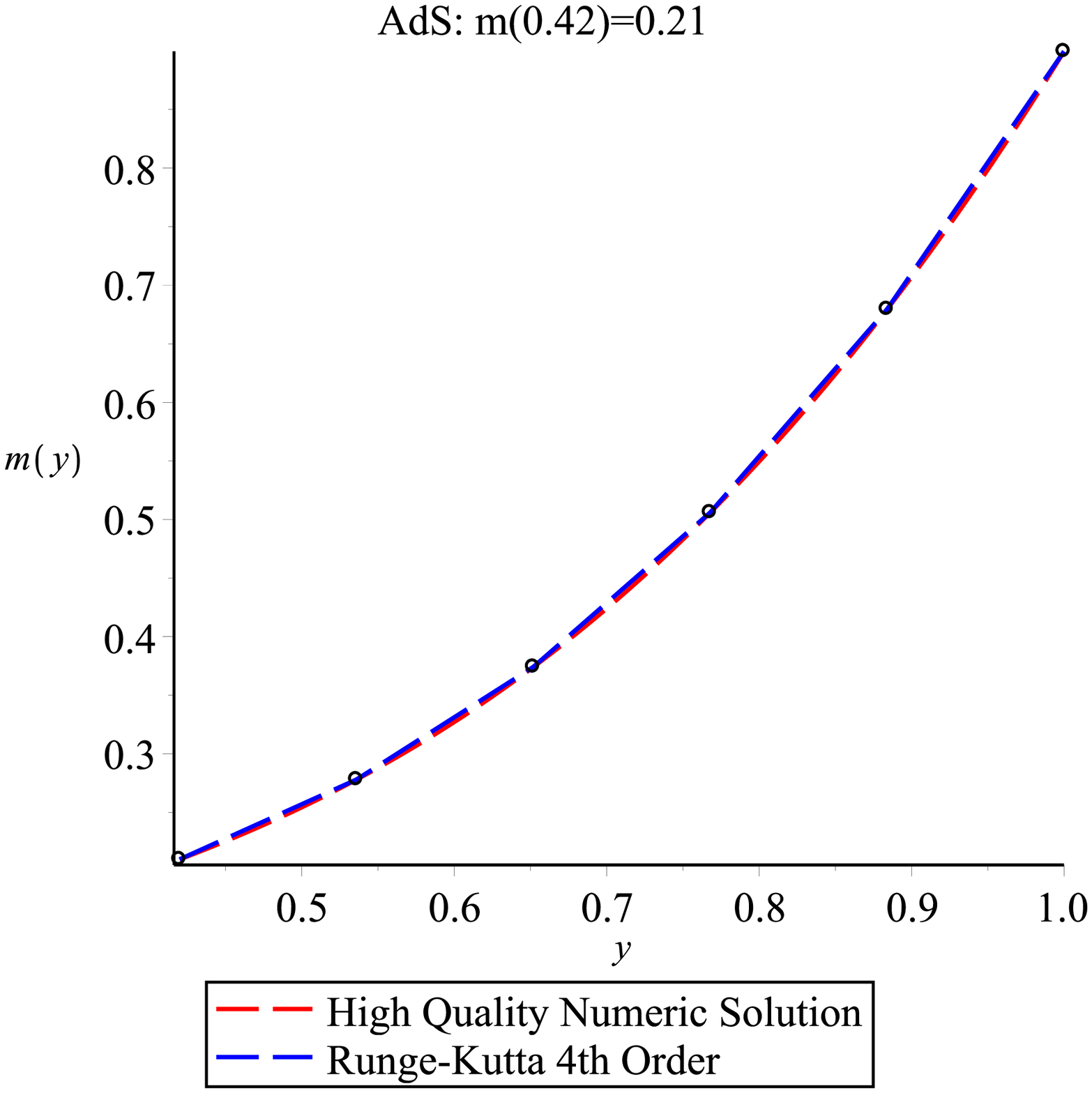}}
\hspace{3mm}\subfigure[{}]{\label{a}
\includegraphics[width=0.3\textwidth]{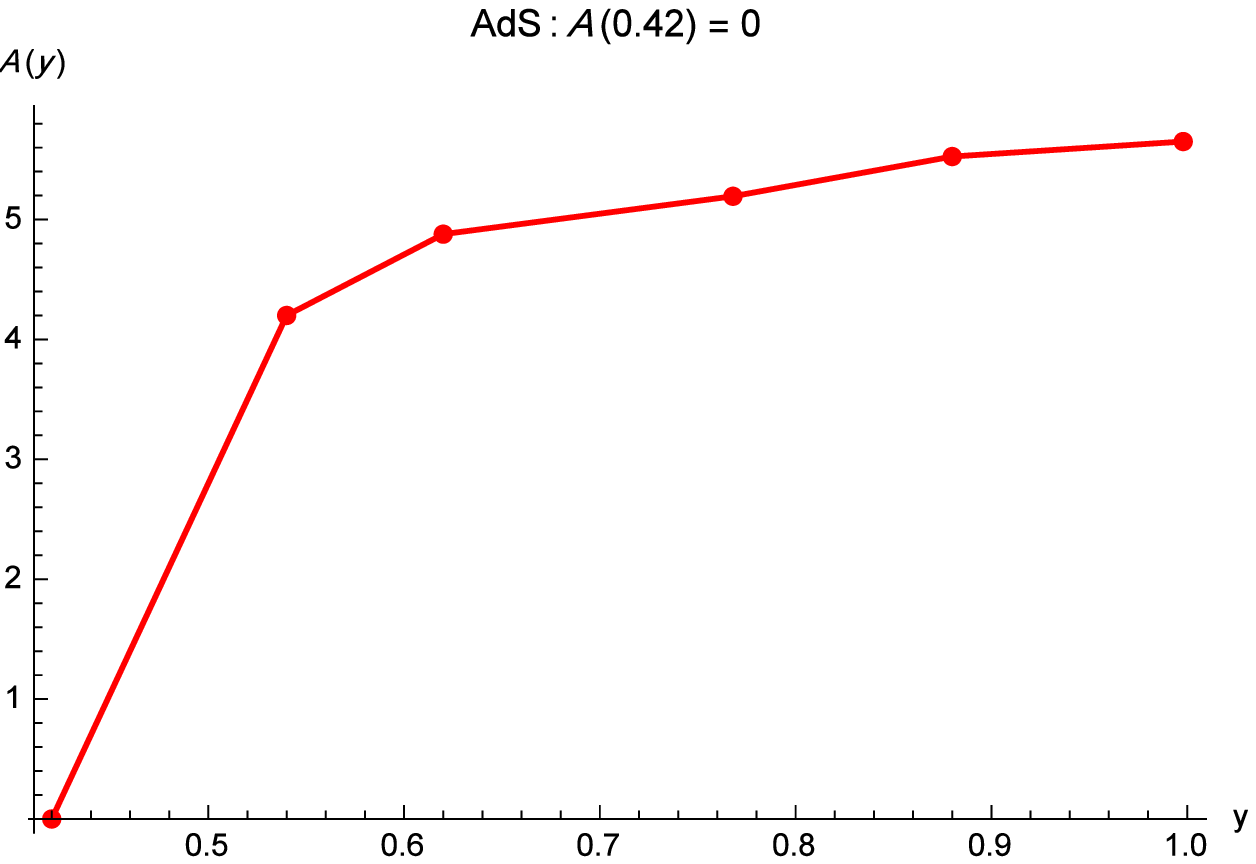}}
\hspace{3mm}\subfigure[{}]{\label{c}
\includegraphics[width=0.3\textwidth]{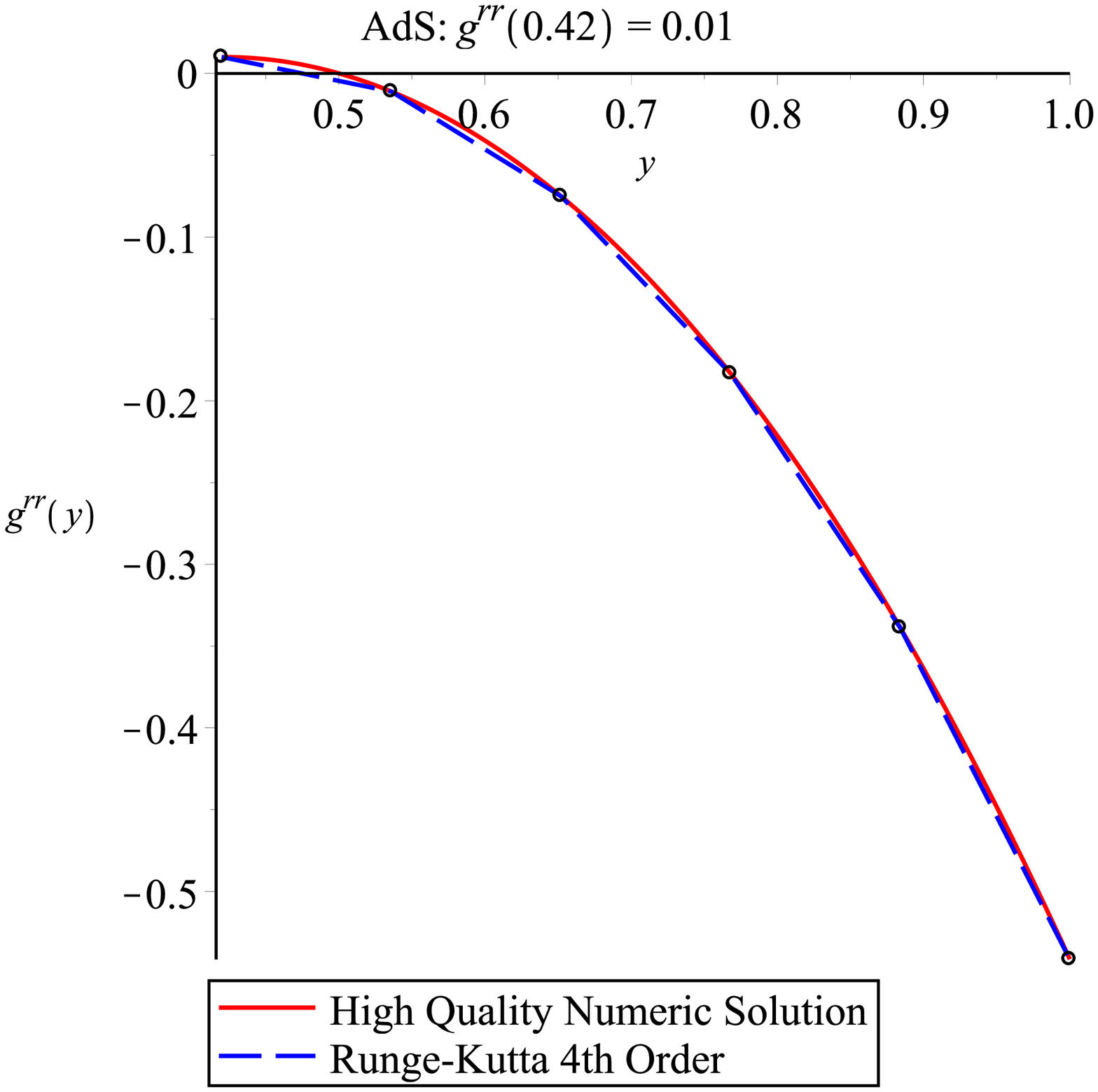}}
\hspace{3mm}\subfigure[{}]{\label{d}
\includegraphics[width=0.3\textwidth]{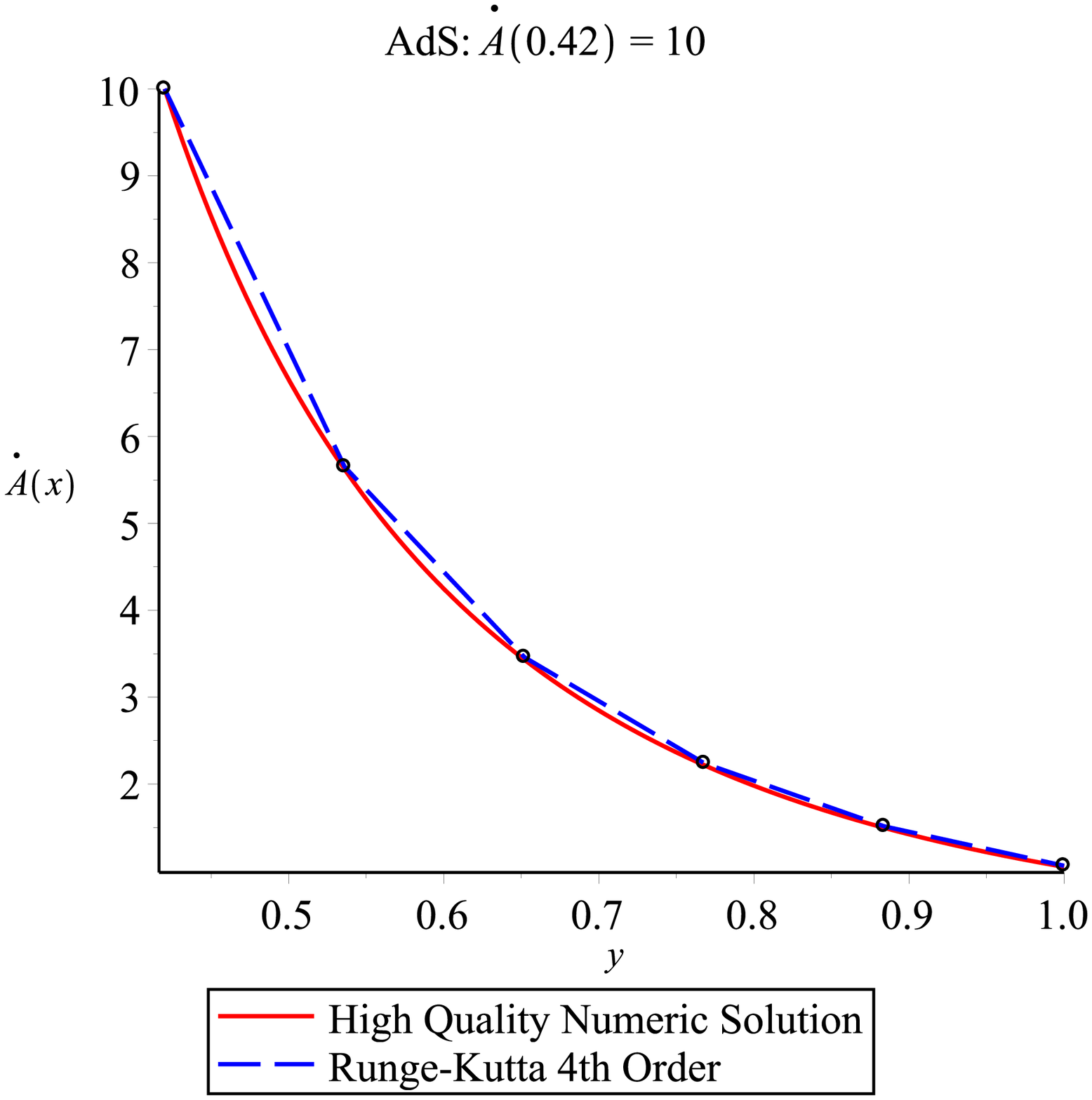}}
\hspace{3mm}\subfigure[{}]{\label{s}
\includegraphics[width=0.3\textwidth]{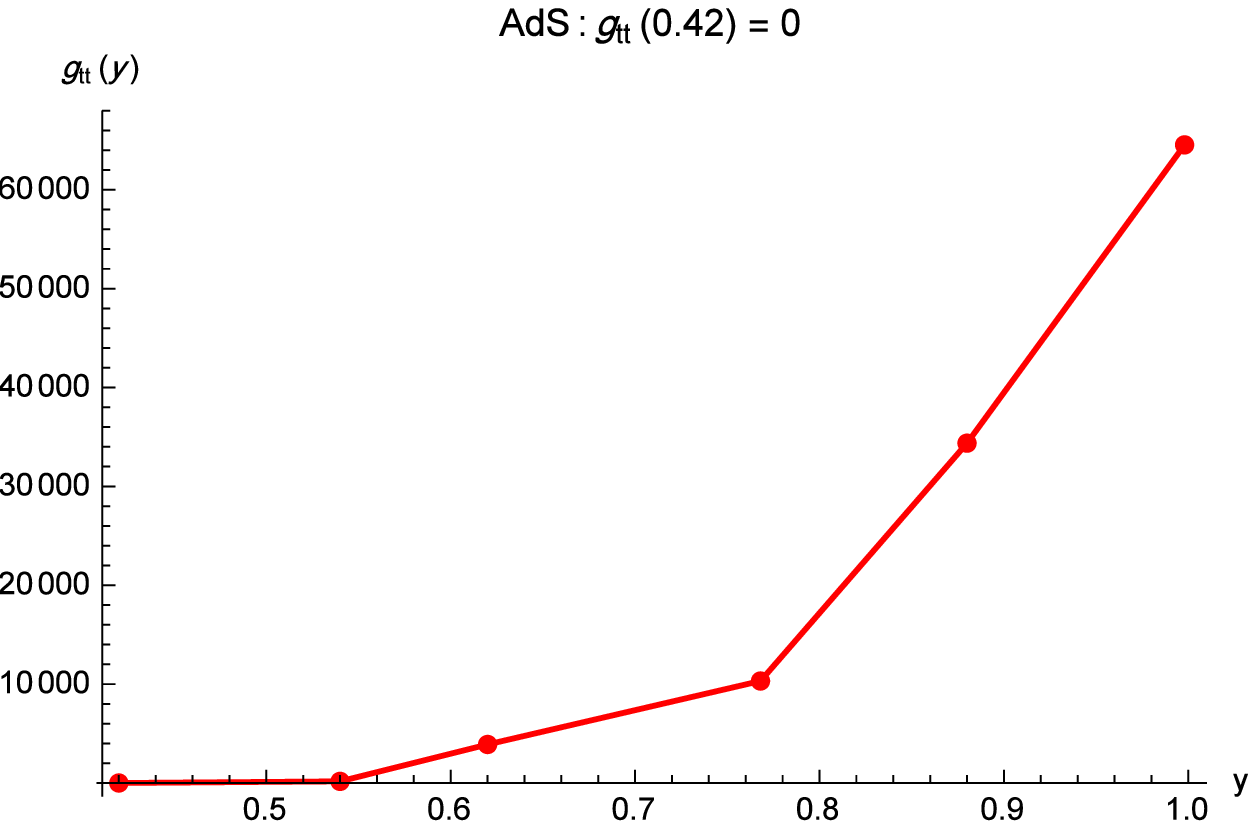}}
\subfigure[{}]{\label{100}
\includegraphics[width=0.3\textwidth]{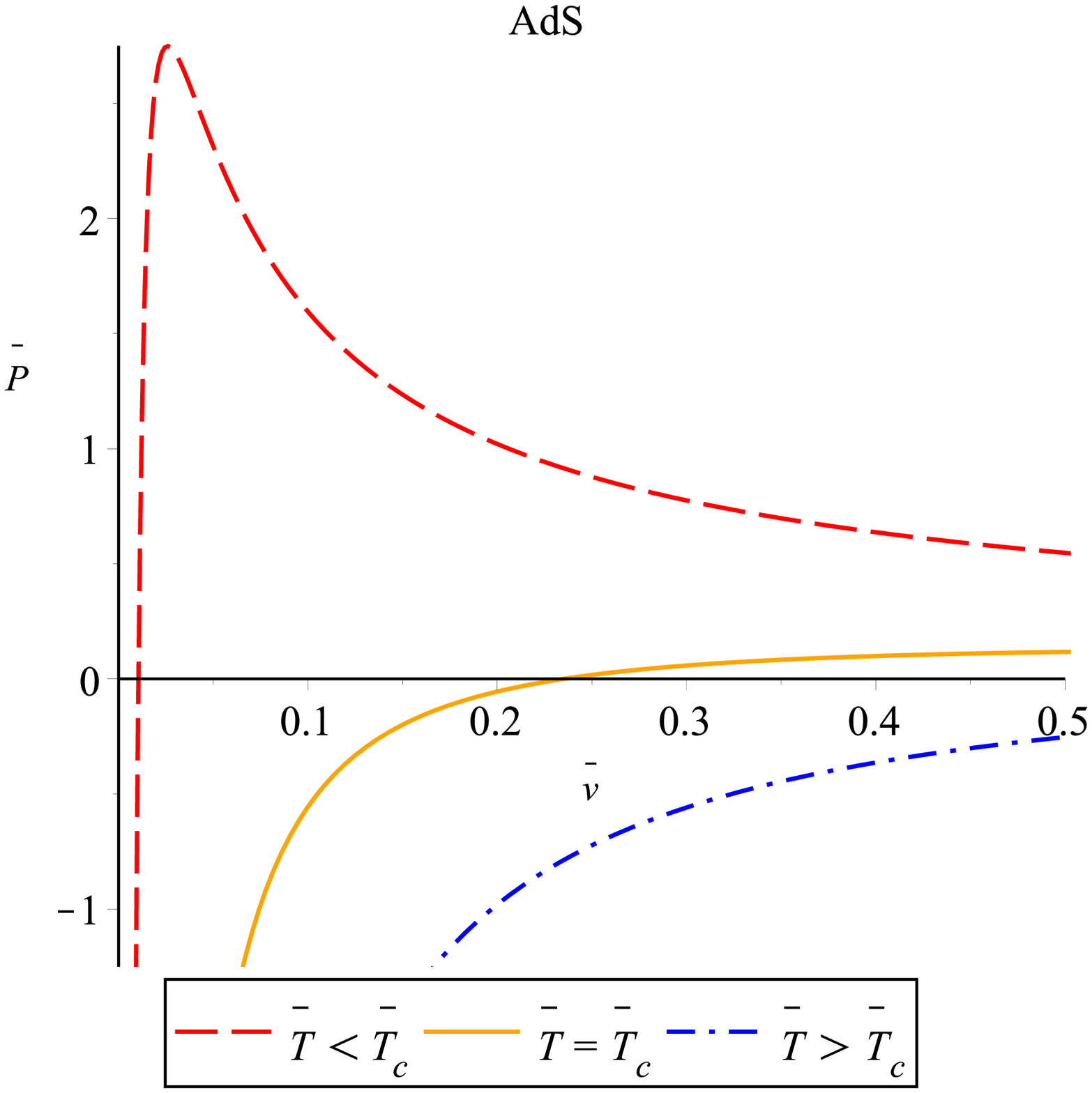}}
\hspace{3mm} \caption{ Numerical solutions are given by diagrams
for AdS sector. Initial values to produce numerical solutions are
shown at top of each diagram.}
\end{figure}

\end{document}